\documentclass[12pt]{article}

\pdfoutput=1

\usepackage{jheppub}
\usepackage{amsmath,amssymb,amsfonts,amscd,mathrsfs}
\usepackage{xcolor}
\definecolor{darkblue}{rgb}{0.1,0.1,.7}

\usepackage{tablefootnote,colortbl}

\usepackage{braket}

\usepackage{cancel}

\usepackage{arydshln}
\usepackage{booktabs,multirow}

\usepackage{tikz-feynman}
 \tikzfeynmanset{compat=1.0.0}
 
\definecolor{myorange}{RGB}{199,146,32}

\definecolor{Gray1}{gray}{0.97}
\definecolor{Gray2}{gray}{0.9}
\definecolor{LightCyan}{rgb}{0.88,1,1}
\definecolor{blu}{rgb}{0,0,1}

\usepackage{ragged2e}
\usepackage{array}
\newcolumntype{L}[1]{>{\raggedright\let\newline\\\arraybackslash\hspace{0pt}}m{#1}}
\newcolumntype{C}[1]{>{\centering\let\newline\\\arraybackslash\hspace{0pt}}m{#1}}
\newcolumntype{R}[1]{>{\raggedleft\let\newline\\\arraybackslash\hspace{0pt}}m{#1}}

\usepackage{dsfont} 
\usepackage{tcolorbox}
\usepackage{booktabs}

\usepackage{titlesec}
\titleformat*{\section}{\large\bfseries}
\titleformat*{\subsection}{\normalsize\bfseries}
\titleformat*{\subsubsection}{\normalsize\it}
\titleformat*{\paragraph}{\normalsize\bfseries}
\titleformat*{\subparagraph}{\normalsize\bfseries}

\def\D {{\Delta_T}}

\newcommand{\reef}[1]{(\ref{#1})}

\def\eps{\epsilon}

\newcommand{\beq}{\begin{equation}} 
\newcommand{\eeq}{\end{equation}}

\def\ge{\geqslant}
\def\le{\leqslant}
\def\geq{\geqslant}
\def\leq{\leqslant}
\newcommand{\diffop}[2]{\ifthenelse{\equal{#2}{1}}{\frac{\mrm{d}}{\mrm{d} #1}}{\frac{\mrm{d}^#2}{\mrm{d} #1^#2}}}

\newcommand{\mrm}[1]{{\mathrm #1}}

\usepackage[normalem]{ulem}

\newcommand{\be}{\begin{equation}}
\newcommand{\ee}{\end{equation}}
\def\bea#1\eea{\begin{align}#1\end{align}}

\newcommand{\matrixel}[3]{\left< #1 \vphantom{#2#3} \right|
	#2 \left| #3 \vphantom{#1#2} \right>} 

\usepackage{accents}
\newlength{\dhatheight}

\numberwithin{equation}{section}
\usepackage{tocloft}
\title{Hamiltonian Truncation with Larger Dimensions}

\author{Joan Elias Mir\'o, }

\affiliation{Abdus Salam International Centre for Theoretical Physics, Strada Costiera 11, 34151, Trieste, Italy}

\author{James Ingoldby}

\date{\today}

\abstract{Hamiltonian Truncation (HT) is a numerical approach for calculating  observables in a Quantum Field Theory non-perturbatively. This approach can be applied to theories constructed by deforming a conformal field theory with a relevant operator of scaling dimension $\Delta$. UV divergences arise when $\Delta$ is larger than half of the spacetime dimension $d$. These divergences can be regulated by HT or by using a more conventional local regulator. In this work we show 
	that  extra UV divergences appear when  using HT rather than a local regulator for  $\Delta \geq d/2+1/4$, revealing 
	a striking breakdown of locality. Our claim is based on the analysis of conformal perturbation theory up to fourth order. As an example we compute the Casimir energy of $d=2$ Minimal Models perturbed by operators whose dimensions take values on either side of the threshold $d/2+1/4$.}

\begin{document}
	
	\maketitle
	\flushbottom

\setcounter{tocdepth}{1}

\newpage
 
\section{Introduction}
\label{innn}

Many quantum field theories (QFTs) can be thought of as points in a renormalization-group (RG) flow from one conformal field theory (CFT) in the ultraviolet to another in the infrared. In principle, these QFTs can be constructed by deforming an ultraviolet CFT using relevant operators. Schematically, the action of these QFTs in $d$ spacetime dimensions is given by
\be
S= S_\text{CFT}+ \sum_i V_{\Delta_i} \, ,  \label{pertFP}
\ee
where $V_{\Delta_i}$ is the integral over spacetime of a local operator in the UV conformal field theory with dimension $0 \leq \Delta_i < d$. When the deformations $V_{\Delta_i}$ are sufficiently weak, we can use perturbation theory to compute physical observables. However, outside of this regime the interpretation of \reef{pertFP} is less clear because we lack a general method to compute observables in theories constructed this way. 

Hamiltonian Truncation (HT) promises to provide a general non-perturbative definition of \reef{pertFP} and allow us to compute its spectrum when the deformation is strong. QFTs constructed as deformations away from a free field CFT are an important special case.
Although mature theoretical tools exist already that are well suited for solving many theories of this kind (such as lattice Monte Carlo), all these tools come with different limitations and sources of systematic error, making it helpful to develop a greater variety of theoretical tools for the analysis of QFTs at strong coupling.

There are already successful demonstrations of HT in the literature for theories in two spacetime dimensions: In the pioneering work of Ref.~\cite{Yurov:1989yu}, HT was used to compute the spectrum of the Lee-Yang two-dimensional CFT perturbed by its only relevant operator (besides the identity). Soon after, Ref.~\cite{Yurov:1990kv} computed the spectrum of the two-dimensional Ising model perturbed by its most relevant operator, and  \cite{Lassig:1990xy} computed the spectrum of the two-dimensional Tricritical Ising model perturbed by each of its relevant primaries.~\footnote{The version of HT used there is often called Truncated Conformal Space Approach (TCSA).} Hamiltonian Truncation has since been applied in a variety of other contexts, for instance to QFTs quantised on the light-cone \cite{Katz:2016hxp,Katz:2013qua,Katz:2014uoa,Fitzpatrick:2018ttk,Anand:2021qnd}, to analyse RG flows on the sphere \cite{Hogervorst:2018otc}, QFTs on an AdS background \cite{Hogervorst:2021spa} and used in combination with the S--matrix bootstrap to analyse form factors \cite{Chen:2021bmm,Chen:2021pgx}. In \cite{Hogervorst:2014rta}, the first HT spectrum computation in $d>2$ spacetime dimensions was carried out for the free-boson CFT perturbed by the  $m^2\phi^2+\lambda \phi^4$ operators. Subsequently, refs.~\cite{Elias-Miro:2020qwz} and \cite{Anand:2020qnp} also found consistent Hamiltonian Truncation spectra for $\phi^4$ theory in $d=3$ spacetime dimensions. 

In this work, we analyse Hamiltonian Truncation in QFTs with $d\geq 2$ that are constructed as deformations away from an ultraviolet CFT. As a preliminary, we describe the Hamiltonian Truncation approach that we use. To calculate the spectrum, we need to regulate infrared divergences. To this end, the CFT is placed on the ``cylinder'' $\mathds{R}\times S_R^{d-1}$, where $R$ is the radius of the $d-1$ dimensional sphere. The Hamiltonian of the QFT thus takes the form
\be
H= H_\text{CFT}+V \ , \quad \text{where} \quad V=g R^{\Delta-d} \int_{S_R^{d-1}}d^{d-1}x\, \phi_\Delta(0,\vec x) \, .  \label{ham1}
\ee
Here, $g$ has been converted into a dimensionless coupling by rescaling it with the cylinder radius $R$, raised to the appropriate power. 
For simplicity, we perturb using a single local operator in this work, but the generalisation to include multiple local operators is straightforward. 
Due to the state-operator correspondence, the CFT Hamiltonian on the cylinder is the dilatation operator   
\be
\langle O_i | H_{CFT}| O_j \rangle= \delta_{ij}\Delta_i /R \, ,
\ee
i.e. $E_i\equiv \Delta_i/R$. We also define the matrix elements $V_{ij}\equiv  R \langle O_i | V | O_j \rangle$. Next, the Hilbert space spanned by the $\{\ket{O_i}\}$ must be truncated. We do this by only including states $\ket{O_i}$ in our truncated basis which have a $\Delta_i$ not exceeding a given Hamiltonian Truncation cutoff $\D$. The approach then proceeds by diagonalizing the finite-dimensional matrix 
\be
\quad \quad \quad \quad \quad \quad \quad  H_{ij}=\Delta_i \delta_{ij} +V_{ij}  \  , \quad \quad   \Delta_i \leq \D  \label{ht1}
\ee
and then extrapolating the spectrum towards the limit where $\D\rightarrow \infty$.
This calculation of the spectrum of $H$ is non-perturbative in $g$, and can  be used to analyse the spectrum of \reef{ham1} even at strong coupling $g\gtrsim 1$.

It would be worthwhile to develop Hamiltonian Truncation into a universal tool that may be used to efficiently compute the spectrum of any theory defined as in \reef{ham1}. To achieve such an objective, various challenges must be addressed. Perhaps the most serious one has to do with UV divergences. 

When the perturbing operator $\phi_\Delta(x)$ has scaling dimension $\Delta \geq d/2$, the spectrum of the truncated Hamiltonian \reef{ht1} fails to converge as $\D\rightarrow \infty$. Such non-convergence can be explained using perturbation theory; the second order perturbative correction to each energy level becomes UV divergent
if $\Delta \geq d/2$. The Hamiltonian Truncation level $\D$ regulates UV divergences in the HT formulation of the theory. A quick solution would be to introduce a counter-term to eliminate the UV sensitivity arising at second order in perturbation theory. This may generically work for operators with dimension $\Delta$ right above $d/2$, but because UV divergences can also appear at higher orders in perturbation theory, we lack a general theory of counter-terms to be used in Hamiltonian Truncation for operators with $\Delta \geq d/2$.

Our goal in this paper is to extend the Hamiltonian Truncation approach, so that it can be applied when the deforming operator takes a larger dimension, i.e. when  $\Delta \geq d/2$. Our strategy is to understand perturbative physics first, and only after aim at non-perturbative computations by diagonalizing \reef{ht1}.

To that end, we analyse the energy levels of \reef{ht1} in perturbation theory. We will regulate our perturbative calculations in two different ways: $(i)$ by imposing a sharp cutoff on the dimension of the QFT Hilbert space, in exactly the same way as in Hamiltonian Truncation and $(ii)$ by using a local position space regulator. We then check whether a perturbative Hamiltonian Truncation computation reproduces the spectrum of a locally-regulated QFT. For simplicity, we focus on the ground state energy. Perturbative corrections to this quantity are observable --  their dependence on the sphere radius $R$ gives rise to a physical Casimir force. 

We show that use of the Hamiltonian Truncation regulator leads to a striking breakdown of locality: we find a regime  $\Delta \geq d/2+1/4$ where the fourth order correction to the ground state energy is finite when a local regulator is used, but leads to a UV divergent result if a HT cutoff is used as a regulator. Due to structural similarities between perturbative corrections to the ground state and excited state energies, this difference between regulators  affects energy differences as well. Our results suggest that adding a local counter-term which removes the UV sensitivity at second order in the perturbation theory is sufficient to make the HT approach well defined, when the deforming operators have scaling dimensions in the range $d/2 \leq \Delta < d/2+1/4$. 

In section~\ref{PC} we review conformal perturbation theory, which will be used throughout the paper. In section~\ref{localHT} we analyse, for general $d$, perturbation theory using both a local position space regulator and a HT cutoff regulator. In section \ref{cptex} and  \ref{CPTHT} we provide explicit examples to support the claims of section~\ref{localHT}. Finally, we conclude in section~\ref{conc} and highlight interesting applications and open problems.

\section{Perturbation Theory}
\label{PC}

At weak coupling, $  g \ll1 $, the spectrum of the Hamiltonian \reef{ham1} can be computed in perturbation theory. In this approach, the energy levels are each expressed as an expansion in powers of the weak coupling $E_i R=\Delta_i + \sum_{n=1}^\infty E^{(n)}_i$. As we are using the Hamiltonian formulation to describe our QFT, we can directly apply Rayleigh--Schr$\ddot{\text{o}}$dinger (RS) perturbation theory to generate this expansion. The first few terms of the RS series are given by 
\bea
E_{i}\, R  &= \Delta_i+
\underbrace{  \phantom{  \bigg|}  V_{ii}      \phantom{  \bigg|}   }_{E_i^{(1)}}+  
\underbrace{   \phantom{  \bigg|}  V_{ik} \frac{1}{\Delta_{ik}} V_{ki}    \phantom{  \bigg|}   }_{E_i^{(2)}}+
\underbrace{   \phantom{  \bigg|} V_{ik} \frac{1}{\Delta_{ik}}  V_{kk^\prime}  \frac{1}{ \Delta_{ik^\prime}}  V_{k^\prime i}   - V_{ii} \, V_{ik}\frac{1}{\Delta_{ik}^2}   V_{ki} }_{E_i^{(3)}}   \nonumber \\[.2cm]
&+   \underbrace{   \phantom{  \bigg|}  \frac{V_{ik}V_{kk^\prime }V_{k^\prime k^{\prime \prime}} V_{k^{\prime \prime}i} }{\Delta_{ik} \Delta_{ik^\prime} \Delta_{ik^{\prime\prime}}}   - 
	E^{(2)}_i \frac{V_{ik} V_{ki}}{\Delta_{ik}^2}  - 2 V_{ii} \frac{V_{ik}V_{kk^\prime}V_{k^\prime i}}{\Delta_{ik}^2\Delta_{ik^\prime}}  - V_{ii}^2 \frac{V_{ik}  V_{ki} }{\Delta_{ik}^3}   }_{E_i^{(4)}}  \,  + \, O(V^5) \label{pert1}
\eea
where $\Delta_{ij}=\Delta_i-\Delta_j$, and a sum over intermediate states $k, k^\prime, k^{\prime\prime}\neq i $ is implicit. In the expansion above, we refer to all the terms entering with a negative sign as subtraction terms. A useful feature of RS perturbation theory is that it makes applying the TCSA as a UV regulator to the $E^{(n)}_i$ straightforward; all that is required is to truncate each of the sums in \reef{pert1} to include only states with $\Delta_k\le\D$.

As our Hamiltonian represents a CFT deformed with a \emph{local} relevant operator, it is also possible to apply conformal perturbation theory. In this formulation, the ground state (g.s.) energy $E_{gs}^{(n)}$ at the $n$th order is expressed as the integrated connected CFT $n$-point function denoted by $\langle \cdots \rangle_c$. For perturbations around a free theory, this corresponds to the usual sum over connected vacuum Feynman diagrams with $n$ vertices. More generally, in conformal perturbation theory, when the deforming operator is primary, it can be expressed as  
\be
E_{gs}^{(n)}  = -( - g)^n     S_{d-1}  /n! \int_{\mathds{R}^d} \prod_{i=1}^{n-1}{d^dx_i}  |x_i|^{\Delta-d} \, \langle  \phi_\Delta(x_1) \cdots \phi_\Delta(x_{n-1}) \phi_\Delta(1)   \rangle_c  \, , \label{pert2}
\ee
where $S_{d-1}=2 \pi^{\frac{d}{2}}/\Gamma(d/2)$ and the spacetime coordinate denoted as 1 represents a unit vector in $\mathds{R}^d$. The conformal perturbation theory for the ground state energy was derived earlier for the $d=2$ case in \cite{Klassen:1990dx}.
The proof of \reef{pert2} follows from considering the exponential decay of the partition function at large (euclidean) time: $E_{gs}(g)= -\lim_{\tau\rightarrow \infty} \frac{1}{\tau}\log \text{Tr}[e^{-\tau H}]$. The form for the integral measure $|x_i|^{\Delta-d}$ is obtained after performing a Weyl coordinate transformation to express the original correlator on the cylinder $\mathds{R} \times  S_{d-1}$ in terms of a correlator on the plane $\mathds{R}^d$. 
We comment now on our notation: we  use $x_i$ and $\phi_\Delta(x_i)$ to denote  coordinates and fields on $\mathds{R}^d$. 
Fields on the cylinder   $\phi_\Delta(\tau,\vec x_i)$ are distinguished by using $(\tau_i, \vec x_i)$ coordinates,  where $\vec x_i$ is a vector in $S^{d-1}$ and $\tau_i$ is the  cylinder time coordinate. 

For convenience, we define a set of constants $c_n$ which characterise corrections to the ground state energy. The first three constants are given by
\be
E_{gs}(g)   =\left[-c_2 \, g^2 /2! + c_3\,  g^3 /3!- c_4\, g^4/4! + O (g^5)\right]  S_{d-1}/R  \, .  \label{genseries}
\ee
The first two constants are then simply equal to integrals of the two and three-point functions
\bea
c_2& =   \int d^dx |x|^{\Delta-d} \langle \phi_\Delta(x)\phi_\Delta(1)\rangle,  \label{c2}\\[.2cm]
c_3 &=\int \prod_{i=1}^2 d^dx_i   |x_i|^{\Delta-d} \langle  \phi_\Delta(x_1)\phi_\Delta(x_2) \phi_\Delta(1) \rangle \, , \label{c3}
\eea
which are necessarily connected because $\langle  \phi_\Delta(x) \rangle=0$. The next constant also receives disconnected contributions:
\begin{multline}
c_4=  \int \prod_{i=1}^3 d^dx_i   \left|x_i\right|^{\Delta-d}    \Big[ 
\langle \phi_{x_1} \phi_{x_2}\phi_{x_3}\phi_{1}\rangle
- \langle \phi_{x_1} \phi_{x_2}\rangle \, \langle\phi_{x_3}\phi_{1}\rangle
\\-\langle \phi_{x_1} \phi_{x_3}\rangle \, \langle\phi_{x_2}\phi_{1}\rangle  
-\langle \phi_{x_1} \phi_{1}\rangle \, \langle\phi_{x_2}\phi_{x_3}\rangle   \Big] \, , \label{c4}
\end{multline}
where we have used  $\phi_{x}\equiv\phi_\Delta(x)$.
This representation of the perturbative coefficients is better suited for regulating the UV divergences using a local position space regulator than the sums over states in  \reef{pert1}.

Energy gaps $E_i-E_{gs}$ can also be expressed in terms of integrals over $n$-point correlation functions. For instance, the energy gap $\delta E_i \equiv (E_i - E_{gs})R$ of the lowest energy state $|\Delta_i\rangle$ overlapping with a primary operator $O_i$ is given by
\be
\delta E_i  =    \Delta_i +   gS_{d-1}  \, C_{OO}^\phi  
-\frac{g^2S_{d-1}}{2}    \int \frac{d^d x}{\left|x \right|^{d-\Delta}}\Big(   \langle O_i(\infty)  \phi_1  \phi_x \,  O_i(0)\rangle_c -  (C_{OO}^\phi)^2  |x|^{-\Delta} \Big)   + O(g^3) \label{pert3} \, ,
\ee 
where the four point function shown above is connected with respect to all four operators, and $O_i(\infty)=\lim_{|s| \rightarrow \infty}|s|^{2\Delta_i}O_i(s)$. By developing perturbation theory for the energy gaps, rather than for the excited state energies directly, the series coefficients become independent of any UV divergences that affect the ground and excited state energies equally, in contrast with alternative approaches \cite{Klassen:1991ze,Giokas:2011ix,Rutter:2018aog}. The derivation of \reef{pert3} is presented in greater detail in appendix \ref{agaps}.

\section{Comparison Between UV Regulators}
\label{localHT}

Next we analyse the UV divergences arising in perturbation theory. We do so by comparing two regulators: a local regulator, and a Hamiltonian Truncation cutoff $\D$ that truncates the Hilbert space on the cylinder to states satisfying $\Delta_i \leq \D$. 


\subsection{Local regulator}
\label{LR}

We refer to a regulator as local if it acts on position space integrals by cutting out infinitesimal regions of the integration domain where the separation between any pair of operators within a correlation function becomes small. 

As an example, we could locally regulate an integral over an $n$--point correlation function in the following way
\be
I_\epsilon = \int \prod_{i=1}^{n-1}d^dx_i \,  |x_i|^{\Delta-d}\,\langle \phi_\Delta(x_1)  \phi_\Delta(x_2)\dots \phi_\Delta(x_n)\rangle \prod_{i<j}\theta\left(|x_i-x_j|-\epsilon\right)\, .
\label{local}
\ee
where for concreteness   \reef{local} is computed in the plane. 
The  Heaviside $\theta$  functions cut out balls of infinitesimal radius $\epsilon$ around each point where an operator is inserted. 
For integrals of products of correlation functions (e.g. the disconnected terms in \reef{c4}) each individual factor is regulated like in \reef{local}.
This type of  regulator is often used in conformal perturbation theory to regulate UV divergencies, see e.g.~\cite{Mussardo:2020rxh}. 
In other contexts, IR divergences may also arise in conformal perturbation theory. However there are no IR divergences for the QFTs defined on $\mathds{R}\times S_{d-1}$ that we consider. In our case,  the $ |x_i|^{\Delta-d}$ factors render the large $|x_i-x_j|\gg 1$ region finite.

There are  detailed properties of the local regulator that we will use  that are not essential for our conclusions in this section. 
For instance,  we could regulate locally by cutting out infinitesimal box shaped regions from the domain of integration, or by cutting out balls with sizes that scale with the position coordinates (for instance by replacing $\epsilon$ in \reef{local} with $\epsilon f(x_i)$ for some smooth function $f(x)$~\footnote{
Regulating integrals over correlation functions on the cylinder $\mathds{R}\times S_{d-1}$ by cutting out $\epsilon$--balls
on the cylinder, and then Weyl transforming the integral  into $\mathds{R}^d$, leads to a local regulator of this type with $f(x)=|x|$ -- which is smooth for all finite values of the cylinder time coordinate $\tau$.}). These regulators break rotational and translational symmetries respectively, and produce different subleading divergences for the same integral, requiring different counterterms for renormalization. 

Our focus in this section however will be on determining whether coefficients in conformal perturbation theory are finite, and for this purpose it will not matter which specific local regulator we use. Determining the finiteness of an integral requires only checking the superficial degree of divergence of the integral at each singularity in this case.

When the two fields in the integrand of \reef{c2} approach one another, the short distance singularity is given by the two point function
\be
\phi_\Delta(x)\phi_\Delta(1) = \frac{1}{|x-1|^{2\Delta}}\, .
\ee
Therefore, by considering the superficial degree of divergence of \reef{c2} around the region $x\rightarrow1$, we see that the integral is not UV divergent for 
\be
d-2\Delta >  0 \, .  \label{sup2}
\ee
Similarly, the three-point function gives the short distance singularities of \reef{c3}
\be
\langle  \phi_\Delta(x_1) \phi_\Delta(x_2)\phi_\Delta(1) \rangle   = \frac{C_{\Delta\Delta}^\Delta}{ |x_1-x_2|^{\Delta} |x_2-1|^\Delta |1-x_1|^\Delta}.
\ee
The leading divergence in the integral comes from the region where both $x_1$ and $x_2$ come close to the unit vector represented by 1. It can be seen that \reef{c3} is finite so long as 
\be
2d- 3\Delta > 0 \, .
\ee

The leading divergence of \reef{c4} cannot be found the same way, as the functional form of the connected four--point function cannot be determined purely from conformal symmetry. Nevertheless, all divergences in this integral come from the regions where separations between $x_1$, $x_2$, $x_3$ and 1 are taken small, and it suffices to know the singularities of the four--point function in these regions.

When one of the separations is taken small with the rest remaining fixed, the singularities of the four--point function are determined by the operator product expansion (OPE). For example, in the vicinity of the singularity at $x_1\rightarrow 1$, we can replace
\be
\phi_\Delta(x_1)\phi_\Delta(1)\sim\frac{1}{|1-x_1|^{2\Delta}}\mathds{1}+\frac{C_{\Delta\Delta}^\omega}{|1-x_1|^{2\Delta-\omega}}\phi_\omega(1)+\dots
\label{opeomega}
\ee
inside the four--point correlator. Here $\phi_{\omega}$ represents any  operator appearing in this expansion. We denote its scaling dimension using $\omega$. Putting \reef{opeomega} back inside \reef{c4}, we find that the singular term proportional to the identity operator cancels with a disconnected term.
Next, by considering the superficial degree of divergence of the integral in this region, we see that the $\phi_\omega$ term contributes no UV divergences so long as
\be
d+\omega-2\Delta>0
\label{opediv}
\ee
for \emph{each} operator $\phi_{\omega}$ appearing in the OPE. 
For all the explicit examples that we discuss in section \ref{cptex}, $\omega\ge\Delta$ and the deformations are relevant $  d>\Delta$ so there will be no UV divergences of this type.

The superficial degree of divergence arising from the region of integration where all coordinate separations are small but hierarchically equal can be determined by considering how the integrand and measure scale under the following conformal transformation:
\be
x_i^\prime= 1+ (x_i-1) \delta \, ,\quad \langle\phi_{x^\prime_1}\phi_{x_2^\prime}\phi_{x_3^\prime}\phi_{1}\rangle_c=
\frac{1}{\delta ^{4\Delta}}\langle\phi_{x_1}\phi_{x_2}\phi_{x_3}\phi_1\rangle_c \, ,\quad \prod_{i=1}^3\frac{d^dx^\prime_i}{|x^\prime_i|^{d-\Delta}}\sim \delta^{3d}\prod_{i=1}^3d^dx_i
\ee
From this scaling analysis we deduce the superficial degree of divergence. 
As a result, there will be no UV divergence from this region, so long as 
\be
3d-4\Delta> 0 \, . \label{sup4}
\ee
Applying a similar scaling argument, we also confirm that no extra UV divergences arise from the region where three of the fields are brought close together, so long as $\phi_\Delta$ is relevant $\Delta<d$.

Using \reef{pert3}, we can check whether the gaps between energies of excited states and the ground state also become UV divergent, up to second order in perturbation theory. The integrand in \reef{pert3} has singularities whenever two of the fields in the correlation function act at the same point, which happens for $x\rightarrow0$, 1 and $\infty$.

In the region where $x$ is close to 1, we can determine the singularities using the OPE shown in \reef{opeomega}. As a result, there will be no UV divergences from here so long as the condition \reef{opediv} is satisfied and $C^\omega_{O_iO_i}\neq0$.

There can be other divergences arising from the $x\rightarrow0$ region of \reef{pert3}, but these should not be interpreted as genuine UV divergences. Instead, they correspond to IR divergences that appear when the chosen interpolating operator $O_i$ overlaps with another excited state with energy lower than $E_i$. More details are provided in Appendix~\ref{agaps}. By analysing the OPE of $\phi_\Delta(x)$ with $O_i(0)$, inserting the result into \reef{pert3} and examining the superficial degree of divergence of the integral, we find that these IR divergences are absent as long as
\be
\omega-\Delta_{O_i}>0,
\ee
for each operator $\phi_\omega$ appearing in the OPE of $\phi_\Delta$ with $O_i$. For the case that $O_i$ itself appears in this OPE, the inequality above is saturated. However, in this case, its term in the OPE produces no IR divergence, because its singularity is canceled by the second term in the integral of \reef{pert3}. The singularities at $x\rightarrow\infty$ are in one--to--one correspondence with those at zero thanks to conformal inversion symmetry, and therefore do not need to be considered separately.

We have shown that a local regulator leaves \reef{c2}--\reef{c4} finite when \reef{sup2}--\reef{sup4} are satisfied, respectively. In the next section we will show that if the Hamiltonian Truncation cutoff regulator is used instead, \reef{c4} diverges in the limit that the cutoff value is taken to infinity even when \reef{sup4} is satisfied, indicating a breakdown of locality.

\subsection{Hamiltonian Truncation regulator}
\label{HT1}

Next we analyse the structure of UV divergences in perturbation theory when the Hamiltonian Truncation regularisation is used. We focus first on the second order perturbation theory correction to the ground state energy. This correction computed using RS perturbation theory is shown in \reef{pert1}. It must also be equivalent to the correction computed using conformal perturbation theory in \reef{c2}. 

To show this equivalence between formulations of perturbation theory explicitly, and to determine how HT regularisation may be applied to \reef{c2}, we rewrite the integrated two point function by Weyl transforming back to the cylinder $\mathds{R}\times S^{d-1}$
\begin{multline}
\int_{-\infty}^\infty d\tau \int_{S^{d-1}} d^{d-1}x\, \langle \phi_\Delta(0,\vec n)\phi_\Delta(\tau,\vec x)\rangle =\\ 2\int_{-\infty}^{0}d\tau\int_{S^{d-1}} d^{d-1}x \sum_{k}e^{\tau\Delta_k}\matrixel{0}{\phi_\Delta(0,\vec n)}{k}\matrixel{k}{\phi_\Delta(0, \vec x)}{0},
\end{multline}
where $\vec n$ represents a point in $S^{d-1}$. We order the correlation function in the cylinder time $\tau$, and insert a complete set of states. 
The factor of $2$ arises because the correlator is invariant under translations in the cylinder time $\tau$.
We extract the $\tau$ dependence by using the time evolution equation $\phi_\Delta(\tau,\vec x)=e^{\tau H}\phi_\Delta(0,\vec x)e^{-\tau H}$ in the interaction picture. Performing the $\tau$ integral yields
\be
 \sum_{k} \frac{2}{\Delta_{k}}\matrixel{0}{\phi_\Delta(0,\vec n)}{k}\int_{S^{d-1}} d^{d-1}x \matrixel{k}{\phi_\Delta(0,\vec x)}{0}.
\label{cyl2}
\ee 
By using \reef{genseries}, and the definition of $V$ in \reef{ham1} it can be seen that this sum representation for the second order correction to the ground state energy equals the RS perturbation theory expression $V_{0k}\Delta_{k}^{-1}V_{k0}$. This tells us that the Hamiltonian Truncation regulator can be applied to the sum \reef{cyl2} by truncating it to include only terms satisfying $\Delta_k\le\D$.

The integrated two-point function on $\mathds{R}^d$ can be expressed as the following sum
\be
\int d^d x | x|^{\Delta-d}\langle \phi_\Delta( 1)\phi_\Delta( x)\rangle  =  \sum_{n=0}^{\infty}\frac{u_n}{2n+\Delta}   \, ,  
\label{ct0}
\ee
where 
\be
u_n =   2S_{d-2}
\sqrt{\pi }\, \Gamma \left(\tfrac{d-1}{2}\right) 
\frac{\Gamma (n+\Delta )}{\Gamma (\Delta )n!} \, 
\frac{\Gamma\left(n+\Delta-\frac{d-2}{2}\right)}{\Gamma \left(\Delta-\frac{d-2}{2} \right)\Gamma \left(n+\frac{d}{2}\right) }
\, 
\label{ht2}.\,
\ee
The sum representation can be derived by expanding the two--point function in the radial coordinate $|x|$. Then by restricting the domain of integration to $|x|\leq1$, using inversion symmetry and integrating each term in the expansion separately, the sum shown in \reef{ct0} is obtained.

The sums \reef{cyl2} and \reef{ct0} must match each other term by term: Expanding the two point function on the plane in powers of the radial coordinate $|x|$ is equivalent to expanding the two point correlator on the cylinder in powers of $e^{\tau}$. Therefore, we should apply the HT regulator to \reef{ct0} by truncating the sum so that $2n+\Delta\le\D$.

To determine whether the second order energy correction is UV divergent when HT is used as the UV regulator, we consider the asymptotic behaviour of the sum in \reef{ct0} for large values of the HT cutoff $\D$. We asymptotically expand the coefficients from \reef{ht2} for large $n$ to yield
\begin{equation}
	u_n\sim2S_{d-2}\sqrt{\pi}\frac{\Gamma(\frac{d-1}{2})}{\Gamma(\Delta)\Gamma(\Delta-\frac{d-2}{2})}n^{2\Delta-d}\left(1+O(1/n)\right).
	\label{unlarge}
\end{equation}
Terms in the sum therefore grow as $\sim n^{2\Delta-d-1}$, and the sum converges provided that $d-2\Delta>0$. This is the same condition for UV finiteness that was found using a local regulator in \reef{sup2}.

By approximating the sum with an integral using the Euler Maclaurin theorem, we find that for $2\Delta-d>0$, the sum grows as
\be
\sum_{n=0}^{2n+\Delta\le\D}\frac{u_n}{2n+\Delta}  \sim S_{d-2} \sqrt{\pi }  \, \frac{ 2^{d-2 \Delta } }{2 \Delta-d} \,   \frac{\Gamma(\frac{d-1}{2})}{\Gamma(\Delta)\Gamma(\Delta-\frac{d-2}{2})} \,  \D^{2\Delta-d}+\dots\, , \label{p1}
\ee
for large $\D$. This result agrees with the computation of \cite{Rutter:2018aog}. Alternatively, this asymptotic behaviour for the divergent sum may be extracted directly from \reef{ct0} using the Hardy--Littlewood Tauberian theorem, which has been used before in a similar context, proving OPE convergence \cite{Pappadopulo:2012jk}.

Next, we discuss the third order correction. The connection between the position space integral, and the RS series representations for this coefficient can be similarly revealed by transforming the integral in \reef{c3} to the cylinder $\mathds{R}\times S^{d-1}$, time ordering and inserting complete sets of states between the operators $\phi_\Delta$
\begin{multline}
c_3 = 
3! \prod_{i=1}^{2}\int_{-\infty\le\tau_1\le\tau_2\le0} d\tau_i\int_{S^{d-1}}d^{d-1}x_i\sum_{k,k^\prime}e^{\tau_1\Delta_{k}}e^{\tau_2(\Delta_{k^\prime}-\Delta_{k})}\\\matrixel{0}{\phi(0,\vec n)}{k^\prime}\matrixel{k^\prime}{\phi(0,\vec x_2)}{k}\matrixel{k}{\phi(0, \vec x_1)}{0}.
\end{multline}
After performing the integrals in $\tau_{1,2}$, we obtain
\be
c_3=3! \prod_{i=1}^{2}\int_{S^{d-1}}d^{d-1}x_i\sum_{k,k^\prime}\matrixel{0}{\phi(0,\vec n)}{k^\prime}\frac{1}{\Delta_{k^\prime}}\matrixel{k^\prime}{\phi(0,\vec x_2)}{k}\frac{1}{\Delta_{k}}\matrixel{k}{\phi(0,\vec x_1)}{0}.
\label{c3sum}
\ee 
Using the definition of $V$ in \reef{ham1}, we see that the previous expression matches the third order energy correction in the RS series, shown in \reef{pert1}. Note that $V_{00}=0$, so the third order subtraction term in the RS series is not present. 
The terms which arise from the matrix multiplications in the RS series expression correspond exactly to the individual terms which appear in the sums over $k$ and $k^\prime$ in \reef{c3sum}. This shows that we would HT regulate the expression by truncating these sums so that $\Delta_k,\Delta_{k^{\prime}}\le\D$.

Now we consider the fourth order correction, whose structure is more intricate than the preceding lower orders. 
In particular there are disconnected terms.
By Weyl transforming to the cylinder and  time ordering the operators in  \reef{c4} we get
\begin{subequations}
\begin{align}
c_4 =  	4!\prod_{i=1}^3\int_\text{t.o.} d\tau_i\int_{S^{d-1}}d^{d-1}&x_i \Big[\, \langle\phi_\Delta(0,\vec{n})\phi_\Delta(\tau_3,\vec x_3)\phi_\Delta(\tau_2,\vec x_2)\phi_\Delta(\tau_1,\vec x_1)\rangle \label{cyl41} \\[-.15cm]
&- \langle\phi_\Delta(0,\vec n)\phi_\Delta(\tau_3,\vec x_3)\rangle\, \langle\phi_\Delta(\tau_2,\vec x_2)\phi_\Delta(\tau_1,\vec x_1)\rangle  \label{cyl42} \\[.2cm]
&-  \langle\phi_\Delta(0,\vec n)\phi_\Delta(\tau_2, \vec x_2)\rangle  \, \langle\phi_\Delta(\tau_3, \vec x_3)\phi_\Delta(\tau_1,\vec x_1) \label{cyl43}\rangle\\[.12cm]
&-  \langle\phi_\Delta(0,\vec n)\phi_\Delta(\tau_1, \vec x_1)\rangle  \, \langle\phi_\Delta(\tau_3, \vec x_3)\phi_\Delta(\tau_2,\vec x_2) \rangle \,  \Big] \label{cyl44}
\end{align}
\end{subequations}
where the fields and correlators are computed in the cylinder and  time ordering (t.o.) means integrating the $\tau$ variables in the  domain  $-\infty\le\tau_i\le\tau_{i+1}\le0 $. 
The last expression for $c_4$ can be matched  to   the fourth order correction of the  RS series
\be
E_{gs}^{(4)} =  -\frac{V_{0k}V_{kk^\prime }V_{k^\prime k^{\prime \prime}} V_{k^{\prime \prime}0} }{\Delta_{k} \Delta_{k^\prime} \Delta_{k^{\prime\prime}}}   + \underbrace{
	\frac{V_{0k} V_{k0}}{\Delta_{k}}  \cdot  \frac{V_{0k} V_{k0}}{\Delta_{k}^2}   }_\text{subtraction terms}   \, , \label{pert4}
\ee
were we have used $V_{00}\propto \langle \phi_\Delta \rangle =0$ to simplify \reef{pert1}, and $\Delta_{0k}=-\Delta_k$.
To perform such matching we  insert the identity representation as a complete set of states $\mathds{1}= \sum_k |k \rangle \langle k|$ between every pair of fields in  the correlators of (\ref{cyl41})-(\ref{cyl44}) and then perform the elementary integrals over the time variables $\tau_i$.

After accounting for the normalization factor $g^4S_{d-1} /( 4!R)$ in 
\reef{genseries}, it can be seen
that the first two terms  \reef{cyl41}-\reef{cyl42} are equal to the   first term in \reef{pert4} while the last two disconnected pieces \reef{cyl43}-\reef{cyl44} are equal to 
the subtraction terms in \reef{pert4}. 
In the next section we show this equivalence explicitly.  
This computation   will allow us to identify the correct implementation of the Hamiltonian Truncation regulator. 

\subsubsection{Analysis of the fouth-order correction  in Hamiltonian Truncation}

We begin by taking just the integral of the four point function \reef{cyl41}.  Inserting complete sets of states between each operator as before and  performing the integrals over the $\tau_i$, we are left with the following expression 
\be
4!\prod_{i=1}^{3}\int_{S^{d-1}}d^{d-1}x_i\sum_{k,k^\prime,k^{\prime\prime}}\matrixel{0}{\phi_{0,\vec n }}{k^{\prime\prime}}\frac{1}{\Delta_{k^{\prime\prime}}}\matrixel{k^{\prime\prime}}{\phi_{0,\vec x_3}}{k^\prime}\frac{1}{\Delta_{k^\prime}}\matrixel{k^\prime}{\phi_{0,\vec x_2}}{k}\frac{1}{\Delta_{k}}\matrixel{k}{\phi_{0,\vec x_1}}{0},
\label{c4sum}
\ee
where we are using the notation $\phi(0,\vec r)\equiv \phi_{0,\vec r}$ for the fields in the cylinder. 
Eq.~\reef{c4sum}  closely resembles the first term appearing in the RS series at fourth order. 
There is however one important difference between the first term of the RS series at fourth order in \reef{pert1} and the HT regulated integral of the four point function in \reef{c4sum}; contributions having $\Delta_{k^\prime}=0$ (which diverge as $1/\Delta_{k^\prime}$) are not counted in the RS series expression, but must be included in \reef{c4sum}, as we are inserting a complete set of states labeled by $k^\prime$, which includes the vacuum state. It turns out however that subtracting  the disconnected  contribution \reef{cyl42} in  $c_4$ removes   these unwanted divergent terms. We now show how this cancellation happens.

Once more we insert the identity resolution between every pair of fields, this time for the time ordered correlators in \reef{cyl42}:
\bea
	4!\prod_{i=1}^{3}\int_\text{t.o.}d\tau_i\int_{S^{d-1}}d^{d-1}x_i\sum_{k,k^{\prime\prime}}e^{\tau_1\Delta_{k}}e^{\tau_2(\epsilon-\Delta_{k})}e^{\tau_3(\Delta_{k^{\prime\prime}}-\epsilon)}& \times\nonumber \\ 
	\matrixel{0}{\phi_{0,\vec n}}{k^{\prime\prime}}\matrixel{k^{\prime\prime}}{\phi_{0,\vec x_3}}{0}  \, &  \, \matrixel{0}{\phi_{0,\vec x_2}}{k}\matrixel{k}{\phi_{0,\vec x_1}}{0},
	\label{d1}
\eea
where we have set the scaling dimension of the ground state to $\epsilon$ to regulate the integral in the infrared. After performing the integrals over the $\tau_i$, we are left with
\be
4!\prod_{i=1}^{3}\int_{S^{d-1}}d^{d-1}x_i\sum_{k,k^{\prime\prime}}\matrixel{0}{\phi_{0,\vec{n}}}{k^{\prime\prime}}\frac{1}{\Delta_{k^{\prime\prime}}}\matrixel{k^{\prime\prime}}{\phi_{0,\vec{x_3}}}{0}\frac{1}{\epsilon}\matrixel{0}{\phi_{0,\vec{x_2}}}{k}\frac{1}{\Delta_{k}}\matrixel{k}{\phi_{0,\vec{x_1}}}{0}.
\label{d0}
\ee
This expression cancels with the $\ket{k^\prime}=\ket{0}$ terms in the sum of \reef{c4sum} in the limit $\epsilon\rightarrow0$, 
after  regulating  the formally divergent expression in  \reef{c4sum} in exactly the same way.
 We may therefore identify the first term of the RS series at fourth order with the integral over the four point function  \reef{cyl41} minus the disconnected piece  \reef{cyl42}.


All in all, by comparing \reef{c4sum} and   \reef{d0} with the RS series we learn that  by inserting 
\be
\sum_{k: \D\geq \Delta_k>0} | k \rangle \langle k  |
\ee
between each of the four operators in \reef{cyl41} we recover the first term of the RS series \reef{pert4}  regulated with Hamiltonian truncation. 

We now focus on the remaining two disconnected contributions at fourth order.
Starting with \reef{cyl43},
we insert  a complete set of states between each pair of fields and perform the $\tau_i$ integrals, leading to
\begin{multline}
	D_2=4!\prod_{i=1}^{3}\int_{S^{d-1}}d^{d-1}x_i\sum_{k,k^\prime}\frac{1}{\Delta_k\Delta_{k^\prime}(\Delta_k+\Delta_{k^{\prime}})}\times\\\matrixel{0}{\phi_{0,\vec{n}}}{k^{\prime}}\matrixel{k^{\prime}}{\phi_{0,\vec{x_2}}}{0}\matrixel{0}{\phi_{0,\vec{x_3}}}{k}\matrixel{k}{\phi_{0,\vec{x_1}}}{0}\, .
	\label{d2}
\end{multline}
By the same token, the third disconnected term \reef{cyl44} can be written as
\begin{multline}
	D_3=4!\prod_{i=1}^{3}\int_{S^{d-1}}d^{d-1}x_i\sum_{k,k^\prime}\frac{1}{\Delta_{k^\prime}^2(\Delta_k+\Delta_{k^{\prime}})}\times\\\matrixel{0}{\phi_{0,\vec{n}}}{k^{\prime}}\matrixel{k^{\prime}}{\phi_{0,\vec{x_1}}}{0}  \, \matrixel{0}{\phi_{0,\vec{x_3}}}{k}\matrixel{k}{\phi_{0,\vec{x_2}}}{0}\, .
	\label{d3}
\end{multline}
We see how to regulate \reef{d2} and \reef{d3} using Hamiltonian Truncation by observing that the sum of those disconnected pieces can be simplified into an expression the same structure as the subtraction term from \reef{pert4} 
\begin{multline}
	D_2+D_3=4!\prod_{i=1}^{3}\int_{S^{d-1}}d^{d-1}x_i\sum_{k,k^\prime}\frac{1}{\Delta_k\Delta_{k^\prime}^2}\times\\\matrixel{0}{\phi_{0,\vec{n}}}{k^{\prime}}\matrixel{k^{\prime}}{\phi_{0,\vec{x_2}}}{0}\matrixel{0}{\phi_{0,\vec{x_3}}}{k}\matrixel{k}{\phi_{0,\vec{x_1}}}{0}\, .
	\label{d23}
\end{multline}
If we truncate both disconnected terms the same way, so that $\Delta_k,\Delta_{k^{\prime}}\le\D$, we can identify \reef{d23} with the subtraction term from \reef{pert4}.

In the RS series, the subtraction term at fourth order has a very similar structure to the second order term. In particular, it depends only on matrix elements of the form $V_{0k}$. Therefore using \reef{ct0}, a general formula for the fourth order subtraction term in HT may be derived, which depends only on the space--time dimensionality $d$, the scaling dimension of the deformation $\Delta$ and the HT cutoff $\D$:
\be
S_A\equiv \frac{4!}{g^4  S_{d-1}}\,\frac{V_{0k} V_{k0}}{\Delta_{k}}\cdot \frac{V_{0k^\prime} V_{k^\prime0}}{\Delta_{k^\prime}^2}\Bigg|_\text{HT}= 3!S_{d-1}\sum_{n_1,n_2=0}^{\substack{2n_1+\Delta\leq  \Delta_T \\ 2n_2+\Delta\leq \Delta_T }}    \frac{u_{n_1}}{(2n_1+\Delta)^2}  \frac{u_{n_2}}{(2n_2+\Delta)}\, , \label{htt}
\ee
where the coefficients $u_i$ were defined in \reef{ht2}. 

\subsubsection{Analysis of large $\D$ behaviour of the fourth order coefficient}

We now consider the large $\D$ behaviour of the fourth order perturbative correction to the ground state energy, when HT is used as the UV regulator. Specifically, we derive the condition for which this coefficient tends to a finite value in the limit $\D\rightarrow\infty$, indicating that it is UV finite. 

The large $\D$ behaviour of the subtraction term in \reef{pert4} can be determined using \reef{htt}. 
We also need to extract the large $\D$ behaviour of the first term of \reef{pert4}, which is given by the four point function \reef{cyl41} minus the disconnected piece in \reef{cyl42}.

\begin{figure}\centering
	A. \;\;\includegraphics[width=0.2\columnwidth]{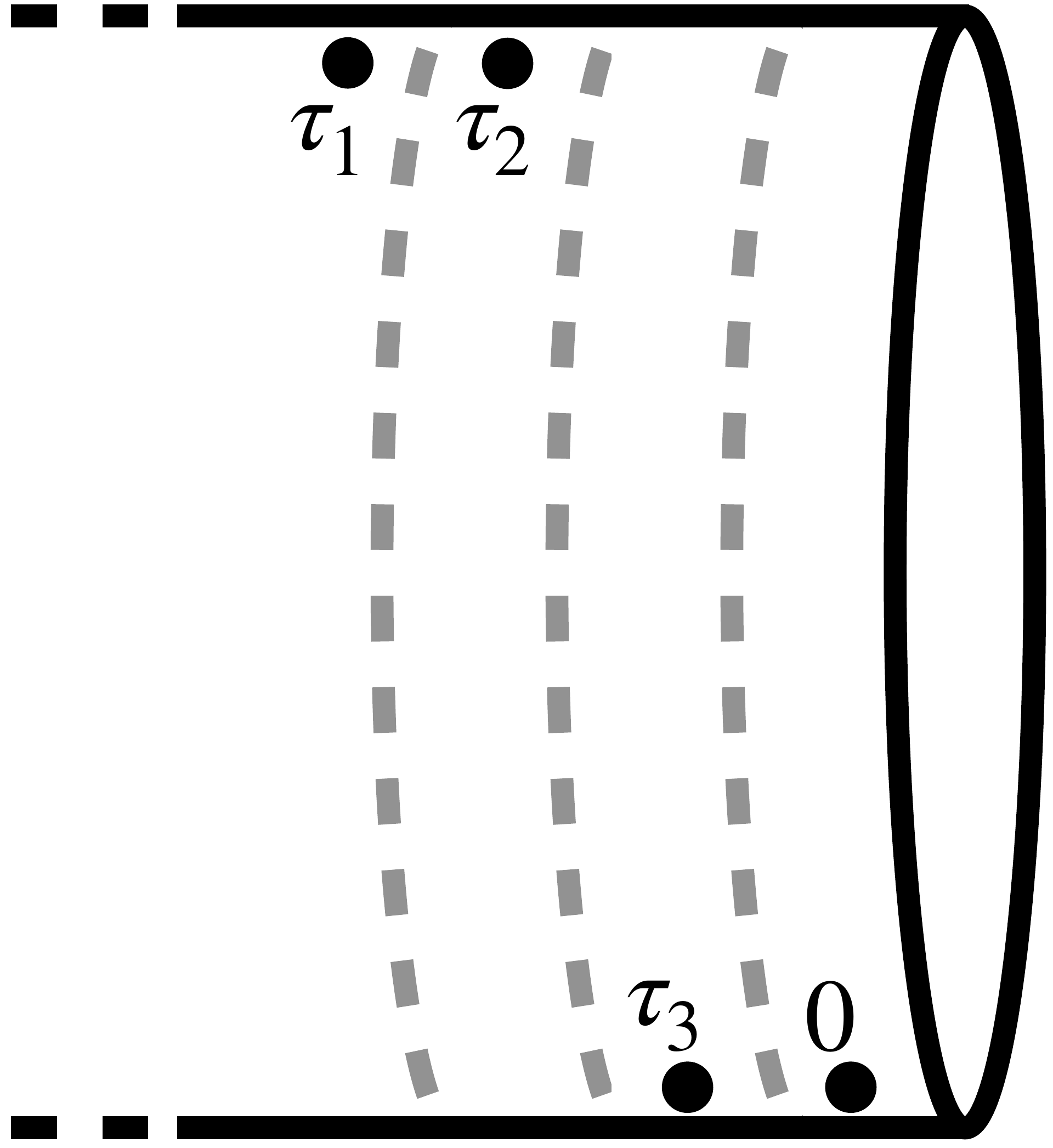}
	\quad \quad 
	B. \;\;	\includegraphics[width=0.2\columnwidth]{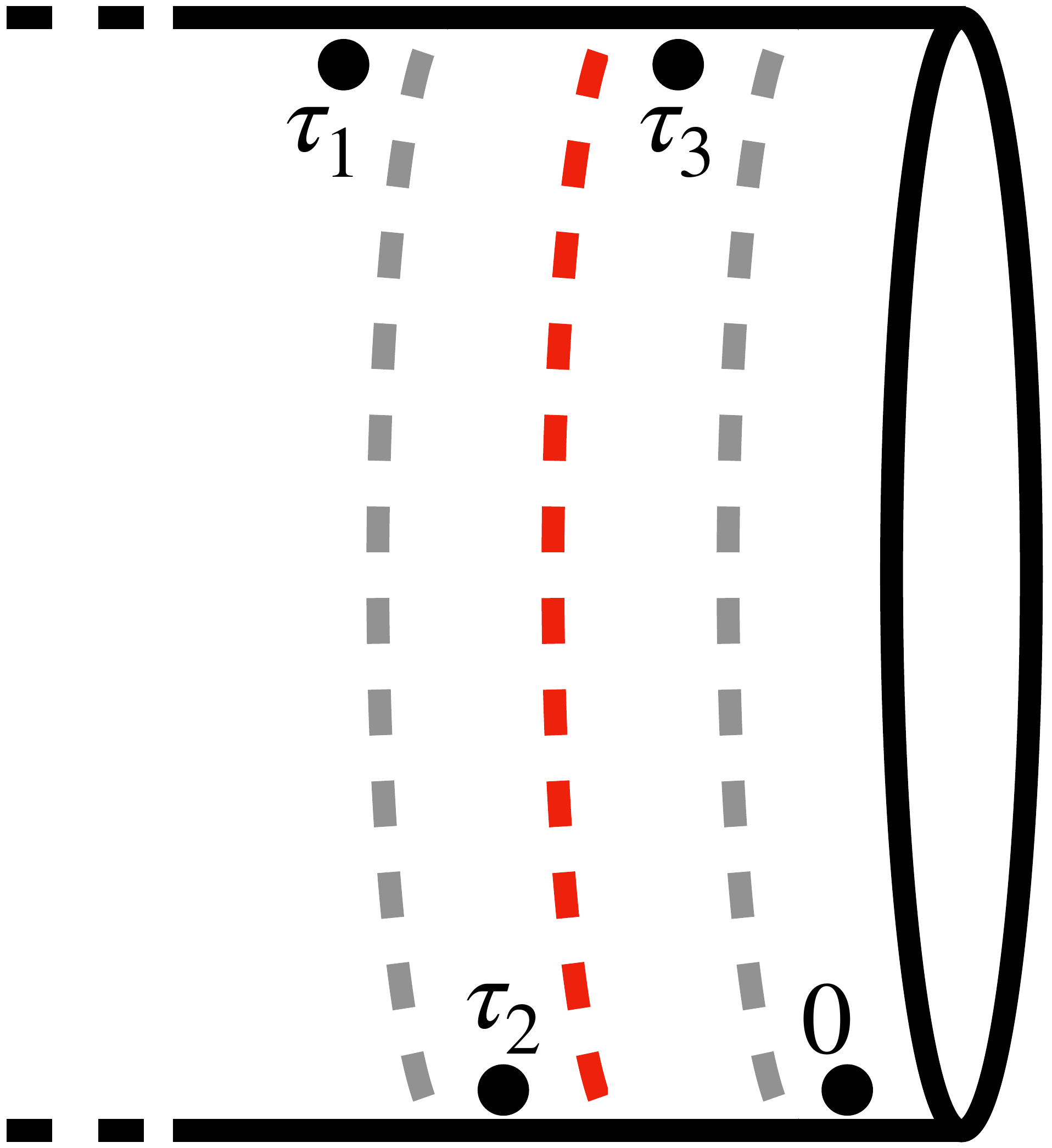}
	\quad\quad 
	C. \;\;\includegraphics[width=0.2\columnwidth]{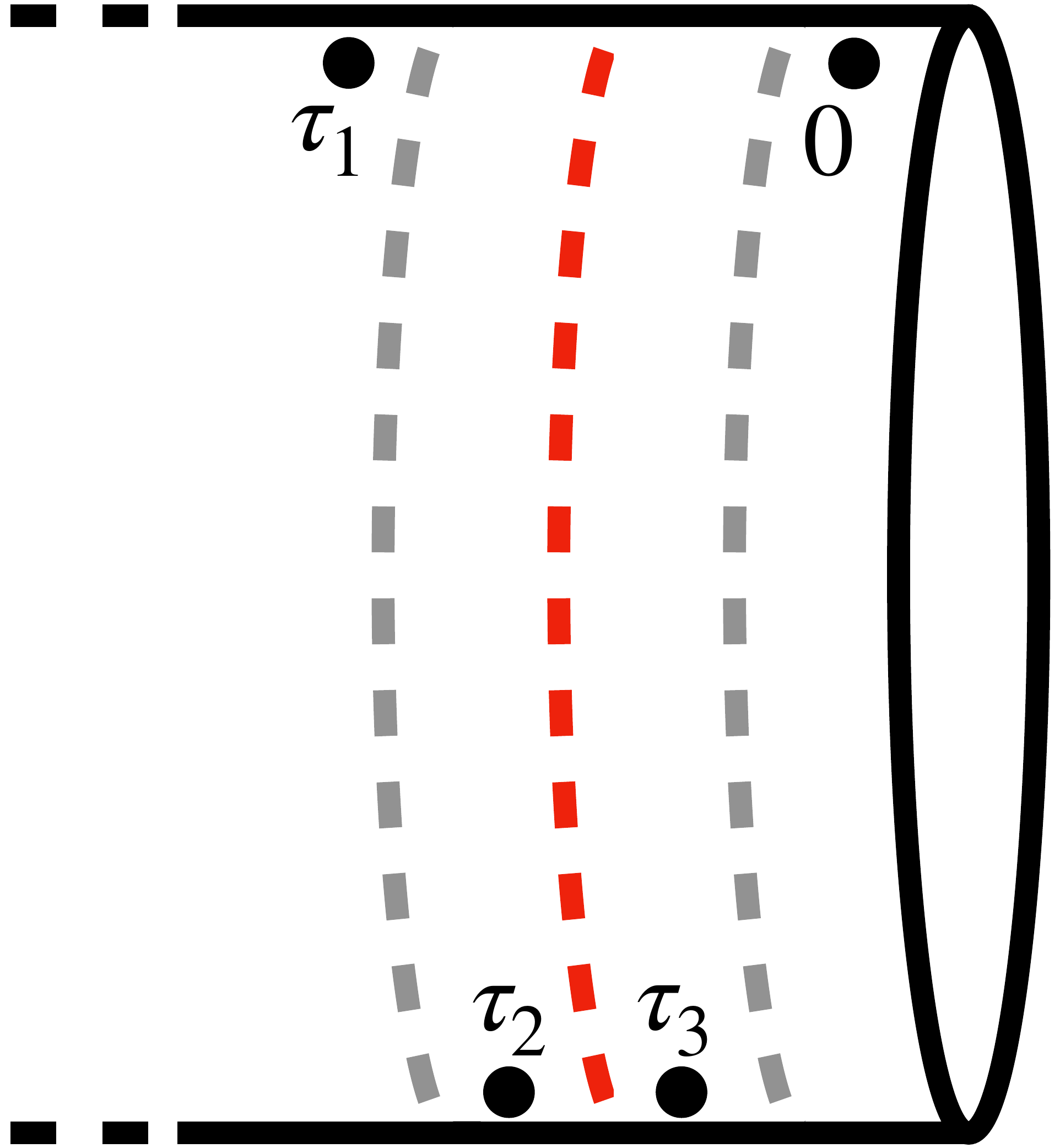}
	\caption{Diagrams representing the three leading singularities of the four point function on the cylinder. The points are all time ordered but have been given angular coordinates so as to position them in  the different singular regions. The dashed lines are placed on surfaces of constant $\tau$, where partial resolutions of the identity $\sum_k^{\Delta_k\le\D}\ket{k}\bra{k}$ are inserted when HT regularisation is used.
	The red dashed lines are discussed in the text. 
	}
	\label{4ptsings}
\end{figure}

Regarding the four point function, the strongest singularities come from three regions where  operators approach one another in pairs. These three singular regions are represented by the diagrams shown in Fig.~\ref{4ptsings}. In each of these regions, use of the OPE shows that the leading singularities of the four point function are the same as those of a couple of two point functions.

The contribution from the singularity which arises when $(\tau_1,x_1)\rightarrow(\tau_2,x_2)$ and $(\tau_3,x_3)\rightarrow(0,n)$ is represented by the diagram in panel A of Fig.~\ref{4ptsings}.  This singularity is exactly canceled when the disconnected piece shown in \reef{d1} is subtracted.

The singularity which arises when $(\tau_1,x_1)\rightarrow(\tau_3,x_3)$ and $(\tau_2,x_2)\rightarrow(0,n)$ is represented by the diagram in panel B of Fig.~\ref{4ptsings}. In this case, when HT is used as the regulator, the singularity of the four point function is no longer canceled when the corresponding integral over the disconnected piece $D_2$  in \reef{d2} is subtracted.

The singularity of the four-point function arising from  this region (shown in figure~\ref{4ptsings},  panel B) can be captured by  the singularity of  
$\int_\text{t.o.}\langle \phi(0,\vec n )\phi(\tau_2,\vec x_2) \rangle\langle \phi(\tau_3,\vec x_3 )\phi(\tau_1,\vec x_1) \rangle$. However, we note  that to regulate the four point function in HT, we must insert three partial resolutions of the identity $\sum_k^{\Delta_k\le\D}\ket{k}\bra{k}$ --  one between each $\phi_\Delta$ operator, as indicated by the dashed lines in panel B of Fig.~\ref{4ptsings}. Accompanying each partial identity insertion is a restriction on the maximum scaling dimension of states which may be evolved across the corresponding line. The strongest condition comes from the middle dashed line, highlighted in red. The states exchanged between $(\tau_1,x_1)$ and $(\tau_3,x_3)$, \emph{and} the states exchanged between $(\tau_2,x_2)$ and $(0,n)$ must both be evolved across the red line simultaneously, leading to the condition $\Delta_k+\Delta_{k^{\prime}}\le\D$. The total scaling dimensions of both sets of exchanged states cannot exceed the cutoff. Integrating this singular term using the HT regulator yields
\begin{multline}
	S_2=4!\prod_{i=1}^{3}\int_{S^{d-1}}d^{d-1}x_i\sum_{k,k^\prime}^{\substack{\Delta_k+\Delta_{k^\prime}\le\D}}\frac{1}{\Delta_k\Delta_{k^\prime}(\Delta_k+\Delta_{k^{\prime}})}\times\\\matrixel{0}{\phi_{0,\vec{n}}}{k^{\prime}}\matrixel{k^{\prime}}{\phi_{0,\vec{x_2}}}{0}    \, \matrixel{0}{\phi_{0,\vec{x_3}}}{k}\matrixel{k}{\phi_{0,\vec{x_1}}}{0}\, .\label{s2}
\end{multline}
Note that the partial sum $S_2$ is cut off differently from $D_2$ in \reef{d2}.

Similarly, the singularity which arises when $(\tau_1,x_1)\rightarrow(0,n)$ and $(\tau_2,x_2)\rightarrow(\tau_3,x_3)$ does not cancel with the disconnected piece $D_3$ in \reef{d3} after integration, when HT is used as a regulator. This singularity is represented by panel C of Fig.~\ref{4ptsings}. 
In this region, we capture the leading singularity from \reef{cyl41} by taking  
$\int_\text{t.o.}\langle \phi(0,\vec n )\phi(\tau_1,\vec x_1) \rangle \, \langle \phi(\tau_3,\vec x_3 )$ $\phi(\tau_2,\vec x_2) \rangle$.
Again for each dashed line in the diagram, a partial resolution of the identity is inserted. Both the set of states exchanged between $(\tau_1,x_1)$ and $(0,n)$ and the states exchanged between  $(\tau_2,x_2)$ and $(\tau_3,x_3)$ must be evolved across the red line, leading to the condition $\Delta_k+\Delta_{k^{\prime}}\le\D$. After integrating the appropriate pair of two point functions and regulating using HT to capture the leading large $\D$ behaviour, we are left with the following expression
\begin{multline}
	S_3=4!\prod_{i=1}^{3}\int_{S^{d-1}}d^{d-1}x_i\sum_{k,k^\prime}^{\substack{\Delta_k+\Delta_{k^{\prime}}\le\D}}\frac{1}{\Delta_{k^\prime}^2(\Delta_k+\Delta_{k^{\prime}})}\times\\\matrixel{0}{\phi_{0,\vec{n}}}{k^{\prime}}\matrixel{k^{\prime}}{\phi_{0,\vec{x_1}}}{0}  \, \matrixel{0}{\phi_{0,\vec{x_3}}}{k}\matrixel{k}{\phi_{0,\vec{x_2}}}{0}\, .\label{s3}
\end{multline}
Individual terms in the partial sum $S_3$ are the same as those of $D_3$ shown in \reef{d3}, but the cutoff of the sum is different.

We emphasise that we did not   deduce the large $\D$ behaviour  
by directly analysing the sums in the first term of  \reef{pert4}.  Instead  we have used the position space representation of the fourth-order correction  to argue that the form of the large $\D$ behaviour 
can be captured by the integrals of two-point functions that we described. To apply the HT regulator to these integrals 
we can compare  the denominators $1/[\Delta_k\Delta_{k^\prime}(\Delta_k+\Delta_{k^{\prime}})]$ from  \reef{s2} (and $1/[\Delta_{k^\prime}^2(\Delta_k+\Delta_{k^{\prime}})]$ from \reef{s3}) with   $1/(\Delta_k\Delta_{k^\prime}\Delta_{k^{\prime \prime}})$ from \reef{pert4}, and deduce  that we should truncate  the states by the condition $\Delta_k+\Delta_{k^{\prime}}\leq \D$.

The sum of terms $S_2$ and $S_3$ determines the leading large $\D$ behaviour  of the first term of \reef{pert4} (up to the normalisation factors in \reef{genseries}).  These terms depend only on matrix elements of the form $V_{0k}$, and therefore a general formula for them may be derived using \reef{ct0}
\be
S_B\equiv S_{2}+S_{3}= 3! S_{d-1} \sum_{n_1,n_2=0}^{2n_1+2n_2+2\Delta\leq\Delta_T}  \frac{u_{n_1}}{(2n_1+\Delta)^2}  \frac{u_{n_2}}{(2n_2+\Delta)}.  \label{locc}
\ee

Next we are ready to evaluate the difference of the two sums $S_A$ (i.e.  the subtraction terms in \reef{pert4} using  HT regularisation) and $S_B$ (captuing the large $\D$ behaviour of the first term in \reef{pert4}). Their difference is not guaranteed to vanish in the large $\Delta_T$ limit. Indeed, we demonstrate this by expanding the difference between these sums for large $\D$. 

We asymptotically expand the summand for large $n_i$ using \reef{unlarge} and approximate the sums with integrals using the Euler--Maclaurin theorem to extract the leading term for large $\D$. We obtain
\be
S_A-S_B= \frac{3!S_{d-1} \;2^{2d-4\Delta}\,S^2_{d-2}\pi\Gamma^2(\frac{d-1}{2})}{\Gamma^2(\Delta)\Gamma^2(\Delta-\frac{d-2}{2})(2\Delta-d-1)}\left\{\frac{\Gamma^2(2\Delta-d)}{\Gamma(4\Delta-2d)}-\frac{1}{2\Delta-d}\right\} \D^{4\Delta-2d-1}+\dots  \label{sab}
\ee
Therefore we find that for
\be
\Delta-d/2-1/4 \geq0 \, , 
\label{redline}
\ee
the difference between the two sums diverges in the $\D\rightarrow\infty$ limit. As the large $\D$ behaviour of the full fourth order correction in HT is captured by this difference between sums, we conclude that the entire Hamiltonian Truncation regulated coefficient will diverge as $\D\rightarrow\infty$ above the threshold \reef{redline}. 



The threshold in Eq.~\reef{redline} also corresponds to the threshold at which subleading divergences arise for the integrals over disconnected parts of the four--point function, such as \reef{cyl43}. A subleading divergence appears if the total superficial degree of divergence of the integral exceeds one. We determine the superficial degree of divergence of the disconnected pieces by separately calculating the superficial degrees of divergence coming from the two regions of the integral where there are singularities. Using the second term of \reef{c4} as an example, we find that these two regions are when $x_1\sim x_2$ and when $x_3\sim 1$. Each region has a superficial degree of divergence of $2\Delta-d$. The total superficial degree of divergence is therefore double this quantity and it exceeds one when threshold \reef{redline} is crossed. For local regulators, subleading divergences are guaranteed to cancel between the integral over the full four--point function and the integrals over disconnected pieces. For the Hamiltonian Truncation regulator, this cancellation no longer happens.

\begin{figure}[t]\centering
	\includegraphics[width=8cm]{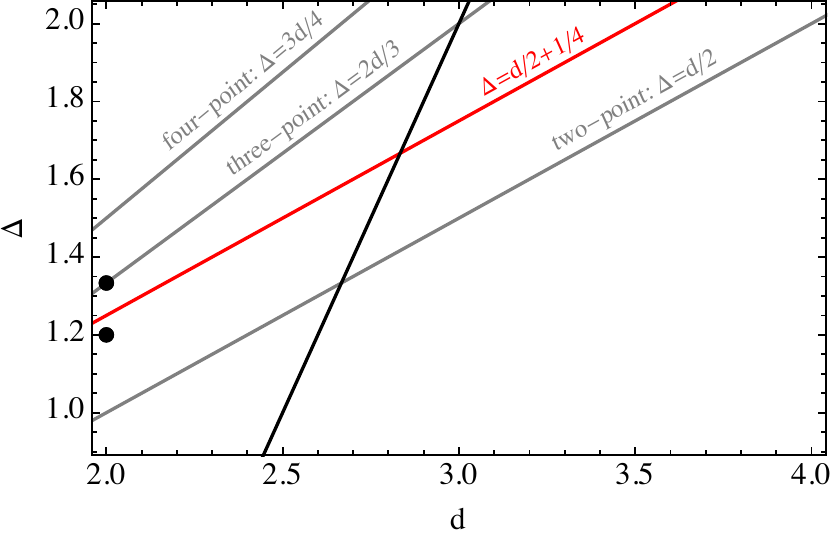}
	\caption{In gray, the values of $\Delta(d)$ for which the integrals of the  two, three and four-point functions in \reef{c2}-\reef{c4} diverge at short distances. In red, the values  for which we expect the four-point function to diverge when using a Hamiltonian Truncation cutoff: $\Delta= d/2+1/4$. This threshold can be probed by, for example: minimal models ${\cal M}_p$, with $p=4,5$, perturbed by the primary operator  $\phi_{13}$, whose dimension is  $\Delta=6/5, 4/3$ respectively (black dots),  and  the free boson perturbed by $\phi^4$, which has dimension $2d-4$ (black line).\label{dimsfig}  } 
\end{figure}

We summarise the analysis of this section in Fig.~\ref{dimsfig}. For perturbing operators with  dimension below each gray line, the corresponding (locally regulated and  connected) $n$-point function is finite. In particular, for perturbing operators with dimension $\Delta \leq 3d/4$ the fourth order corrections to the Casimir energy is well defined, in the  local regularisation scheme. For perturbing operators with dimension above the red line the integral of the corresponding connected four-point function diverges when the Hamiltonian Truncation regulator is used.

This extra divergence in Hamiltonian Truncation  that we discussed was found earlier in the context of $\phi^4$ in three spacetime dimensions~\cite{Elias-Miro:2020qwz}.
There the HT computations were based on perturbing the  free  massive scalar  by the  $\phi^4$ operator.
 The HT regulator of this theory spoils an important cancelation between disconnected Feynman diagrams entering in the perturbative calculation of the vacuum energy (see in particular Eq.~(5.10) of ref.~\cite{Elias-Miro:2020qwz}). 
 This finding is  consistent with our bound because $\Delta_{\phi^4}= 2(d-2)>d/2+1/4$ for $d=3$.~\footnote{The mass gap of this theory  is only logarithmically divergent, which  means that there are no sub-leading divergences. Therefore it is possible  to get nice data for  the mass gap  at strong coupling using HT.}
 
In the following sections  we show the effect of this new threshold (red line) in two more example theories. In particular,
we will calculate the fourth order coefficient $c_4$  in  theories sitting on either side of the red line using both a local and the Hamiltonian Truncation regulator. 
We show  that the coefficient is finite for the theory below the red line if either regulator is used. We also show that for the theory above the line, the Hamiltonian Truncation regulated coefficient diverges as $\D\rightarrow\infty$, whereas the locally regulated one does not.
The two example theories that we will use to test our condition \reef{redline} are the diagonal minimal models~\cite{Belavin:1984vu} ${\cal M}_p$, for $p=4,5$. We perturb these two--dimensional CFTs using the $\phi_{13}$ operator, whose scaling dimensions are $\Delta=6/5, 4/3$, respectively. 


\section{Conformal Perturbation Theory Examples}
\label{cptex}

In this section, we examine conformal perturbation theory for the minimal model CFTs ${\cal M}_4$ and ${\cal M}_5$ perturbed with a relevant operator. The central charge of  the ${\cal M}_p$ CFT is $c_p=1-6/(p(p+1))$, where $p=2,3,4,\dots$.
These CFTs describe the critical points of the Landau-Ginzburg theory $S[\varphi]\sim\int ( \partial \varphi)^2+ \varphi^{2(p-1)}$, and for instance $p=3$ describes the critical point of the Ising model. We consider the vacuum energy of the theory on the cylinder defined by the euclidean action
\be
S=S_{{\cal M}_p}+ g R^{\Delta-2} \int \phi_{13}(z, \bar z) \,   d^2z   \, ,  
\label{model1}
\ee 
where $g$ is dimensionless and the dimension of the operator $\phi_{13}$ is $\Delta=2-4/(p+1)$. Primary operators in minimal models may be organised into a Kac table and labeled using a pair of integers corresponding to their position in the table. The subindex $(1,3)$ refers here to the choice made for the relevant perturbing operator. A more detailed introduction to minimal models is provided, for example, in \cite{Mussardo:2020rxh}. We have used the holomorphic coordinate $z=x^1+i x^2$ in \reef{model1}, with its complex conjugate given by $\bar z$. Subsequently, we shall use $\{ x^\mu_j\}$ to denote cartesian, and $\{z_j\}$ to denote holomorphic coordinates.~\footnote{The $\phi_{13}$ perturbation of minimal models preserves the integrability of the CFT~\cite{Zamolodchikov:1991vx,Fendley:1993xa}. This feature will not play an essential role in our discussion. }

In order to compute perturbative corrections to the ground state energy using \reef{c2}-\reef{c4}, we first calculate the two, three, and four point functions of $\phi_{13}$ in the two minimal models. Conformal symmetry constrains the two and three point functions to take the forms below:
\begin{gather}
\langle\phi_{13}(z_1) \phi_{13}(z_2) \rangle = \frac{1}{|z_1-z_2|^{2\Delta}}, \nonumber \\ \langle\phi_{13}(z_1) \phi_{13}(z_2)\phi_{13}(z_3) \rangle = \frac{C_{}}{|z_1-z_2|^\Delta |z_2-z_3|^\Delta |z_3-z_2|^\Delta}  \, . \label{twothree}
\end{gather}
 The structure constant is given by~\cite{Dotsenko:1984ad}
 \be
 C\equiv C_{(1,3)(1,3)}^{(1,3)}=  \begin{cases}
      \frac{2}{3} \sqrt{\frac{\Gamma \left(\frac{4}{5}\right) \Gamma \left(\frac{2}{5}\right)^3}{\Gamma \left(\frac{1}{5}\right) \Gamma \left(\frac{3}{5}\right)^3}}   & \text{if}\ p=4 \\[15pt]
  \sqrt{\frac{8}{9}}   & \text{if}\ p=5
    \end{cases} \quad . 
 \ee
The four point functions for these theories can be computed exactly, using the coulomb gas formalism~\cite{Dotsenko:1984nm}. They take the general form
\be
\langle  \phi_{13}(z_1) \cdots  \phi_{13}(z_4) \rangle = 
  \frac{|z_{13}z_{24}|^{-4-4\Delta}}{|z_{12}z_{23}z_{34}z_{41}|^{-2-\Delta}}
  \,  \left( \alpha_1 |J_1(\eta)|^2 + \alpha_2 |J_2(\eta)|^2 + \alpha_3 |J_3(\eta)|^2\right)  \, ,  \label{tri4pt}
\ee
where $z_{ij}= z_i-z_j$, and the quantity $\eta=(z_{12}z_{34})/(z_{13}z_{24})$ is the conformal cross ratio. 

For $p=4$, the functions $J_i$ are  
\be
J_i =   \frac{1+\eta(\eta-1)}{\left(\eta(1-\eta)\right)^{5/2}}  \, \big[u_i \,  P_{-1/5}^{3/5}(-1+2\eta) + v_i \, Q_{-1/5}^{3/5}(-1+2\eta) \big] \, ,  \label{p4I}
\ee
where $P$ and $Q$ are the Legendre functions, and $\{\alpha_i, u_i,v_i\}$ are constants which we give in appendix~\ref{afourpt}. 
 While for $p=5$, one has
\bea
J_1&=\frac{3 \left((1-\eta)^{2/3}-1\right)+\eta \left(-2 (1-\eta)^{2/3}+\left(3 (1-\eta)^{2/3}-4\right) \eta+4\right)}{\eta^3 (\eta-1)^3 }\, , \nonumber \\
 J_2 &=  \frac{(4-3 \eta) \eta-4}{(\eta-1)^3 \eta^{7/3}}\, ,  \nonumber\\
 J_3 &= \frac{\left(3 (1-\eta)^{2/3}+4\right) \eta^2-2 \left((1-\eta)^{2/3}+2\right) \eta+3 \left((1-\eta)^{2/3}+1\right)}{ \eta^3(\eta-1)^3} \, .  \label{p54pt}
\eea
We review the formalism we used to derive \reef{tri4pt}-\reef{p54pt} in greater detail in section~\ref{CG}.
To our knowledge  these four-point functions   have not been computed explicitly before in the literature.

\subsubsection*{The vacuum energy}

Now that the correlation functions \reef{twothree} and \reef{tri4pt} are known, we compute perturbative corrections to the ground state energy using \reef{c2}-\reef{c4}. We first consider the  Tricritial Ising Model ${\cal M}_4$ perturbed by $\phi_{13}$. The second order correction is not finite because \reef{sup2} is not satisfied for $\Delta=6/5$. The higher order corrections are finite though. The third order correction can be computed by analytically performing the integral \reef{c3} of the three-point function in \reef{twothree}  
\be
\frac{c_3}{(2\pi)^3} = \frac{C}{4 \pi^2}  \, 
  \frac{\Gamma \left(\frac{\Delta }{4}\right)^2}{\Gamma \left(1-\frac{\Delta }{4}\right)^2}    
  \frac{\Gamma (-\frac{\Delta}{2} )}{\Gamma (1-\frac{\Delta}{2} )}
  \frac{\Gamma (\frac{\Delta}{2} )}{\Gamma (\frac{\Delta}{2} +1)}
  \frac{\Gamma \left(\frac{\Delta }{4}+1\right)}{\Gamma \left(-\frac{\Delta }{4}\right)}
  \frac{\Gamma \left(1-\frac{3 \Delta }{4}\right)}{\Gamma \left(\frac{3 \Delta }{4}\right)}  
   \simeq 1.32 \, ,   \label{simeq1}
\ee
where we introduced a convenient $1/(2\pi)^3$ normalisation. 
We estimate the fourth order correction by computing the integral of the four-point function numerically, using \reef{tri4pt} and \reef{twothree}. We find
\be
\quad \quad\quad \quad \quad \frac{c_4}{4!(2\pi)^3} \simeq -0.18\pm 0.02 \,   \quad \quad \quad (p=4)   \label{simeq2}
\ee
As we shall see in section \ref{CPTHT}, finding the numerical integral to this level of precision is sufficient for checking agreement between the results of local and Hamiltonian Truncation regularisation. The connected four--point function has integrable singularities within the domain of integration, which make significantly increasing the precision of our estimate nontrivial.
 
Next, we consider the ground state energy of the ${\cal M}_5$ minimal model perturbed by $\phi_{13}$. We note that the two and three point functions are UV divergent. 
To obtain the fourth order correction, we perform the integral in \reef{c4} using \reef{p54pt} and find
 \be
\quad \quad\quad \quad \quad \frac{c_4}{4!(2\pi)^3} \simeq 1.0\pm 0.1 \,   \quad \quad \quad (p=5) \label{simeq3}
\ee
In section \ref{CPTHT}, we will compare the results in \reef{simeq1}--\reef{simeq3} to the corresponding values for these quantities regulated using the Hamiltonian Truncation.

\subsection{Four--Point Function Particulars}
\label{CG}

In this section we describe the derivation of the equations \reef{tri4pt}-\reef{p54pt} for the correlator $\langle \phi_{13}\phi_{13}\phi_{13}\phi_{13} \rangle$ in greater depth. We start with a brief review of the parts of the coulomb gas formalism that we will need, and then guide the reader through the computation of the four--point function of interest. 
 
The Dotsenko and Fateev~\cite{Dotsenko:1984nm} coulomb gas formalism can be used to provide an integral representation for the $n$--point correlation functions of the minimal models. 
The starting point is the action of a free boson coupled to the scalar curvature $R(g)$,
 \be
 S=\frac{1}{8\pi}\int d^2 x \sqrt{g} [\partial_\mu \varphi \partial^\mu \varphi+  i 2 \sqrt{2}\alpha_0 \varphi R(g)] \, .  \label{act}
 \ee
 The coupling $\alpha_0\varphi R(g)$ reduces the central charge of the free theory from one to  $1-24\alpha_0^2$.
 The action \reef{act} is not real, but it leads to a unitary conformal theory for values of the central charge $c_p=1-24\alpha_0^2$ which coincide with the central charges of the unitary minimal models ${\cal M}_p$, namely for $\alpha_0^2=1/[4 p(p+1)]$. 
 The vertex operators of this theory 
 \be
 V_{\alpha}(z,\bar z)=e^{i \sqrt{2}\alpha \varphi(z,\bar z)} \, ,
 \ee
 are conformal primaries with $U(1)$ charge $\sqrt{2}\alpha$ and scaling dimension $\Delta_\alpha= 2\alpha( \alpha-2\alpha_0)$, which is invariant under the exchange $\alpha\rightarrow 2\alpha_0-\alpha$.
Correlation functions of vertex operators are given by
\be
\langle V_{\alpha_1}\cdots V_{\alpha_n} \rangle = \prod_{i<j} |z_i-z_j|^{4\alpha_i \alpha_j} \, . \label{corr}
\ee
In the free theory $\alpha_0=0$, the conservation of the $U(1)$ charge associated with the Noether current ${\cal J}(z)=i\partial\varphi$ 
imposes the neutrality condition $\sum_i \alpha_i=0$, which must be satisfied otherwise \reef{corr} vanishes. However, the coupling of the scalar to the topological term $R(g)$, changes the Ward identity of the $U(1)$ symmetry with the net effect of imposing the modified condition $\sum_i \alpha_i=2 \alpha_0$ on \reef{corr}.
 
In the coulomb gas formalism, the conformal fields $\phi_{r,s}$'s of the minimal models are identified with vertex operators  $V_{\alpha_{rs}}$ having the same conformal dimensions. Here the labels $\{r,s\}$ refer to coordinates specifying the position of the operator within the Kac table. 
In the correlation functions,  the  conformal field $\phi_{r,s}$ can be represented by either $V_{\alpha_{rs}}$ or $V_{2\alpha_0-\alpha_{rs}}$ due to the invariance of the scaling dimension under the exchange $\alpha_{rs}\rightarrow 2\alpha_0-\alpha_{rs}$.
However, because 
the non-vanishing correlation functions $\langle V_{\alpha_1}\cdots V_{\alpha_n} \rangle$ must satisfy the selection rule $\sum_i \alpha_i=2\alpha_0$,
it is not possible to write down a  four-point function using only $V_{\alpha}$ and $V_{2\alpha-\alpha_0}$ such that the total charge adds up to $2\alpha_0$. This poses a problem if we are to identify $\langle \phi_{r_1,s_2}\cdots   \phi_{r_n,s_n} \rangle $ with non-vanishing correlation functions of vertex operators. 
As shown by Dotsenko and Fateev~\cite{Dotsenko:1984nm}, this problem can be solved by introducing screening charge operators
\be
Q_{\pm}= \oint V_{\alpha_\pm}(z)dz
\ee 
with $U(1)$ charge $\alpha_\pm\equiv \alpha_0\pm \sqrt{\alpha_0^2+1}$. Here, we have used the holomorphic notation so that $V_{r,s}(z)$ should be understood as $V_{r,s}(z,\bar z)\equiv V_{r,s}(z)\otimes\bar{V}_{r,s}(\bar z)$. 
The  conformal dimension of $V_{\alpha_\pm}$ is  $\alpha_\pm(\alpha_\pm-2\alpha_0)=1$, and therefore $Q_\pm$ have zero conformal dimension because $V_{\alpha_\pm}(z) dz$  is a singlet under  conformal transformations.

To get the correlation functions of $\phi_{13}$, we compute correlators of the form \newline $\langle  V_{\alpha_1} \cdots  V_{\alpha_1} Q_+^n Q_-^m   \rangle $, where  the vertex operators $V_{\alpha_i}$ are either $V_{\alpha_{1,3}}$ or $V_{2\alpha_0-\alpha_{1,3}}$. The integers $\{m,n\}$ are fixed in order to set the total charge of the $n$-point function to $2\alpha_0$, so that the correlator does not vanish. Typically, the fewer the screening charges present, the simpler it is to compute the  correlator. The simplest representation of the four--point function is given by 
 \bea
  \langle  \phi_{13}(0)  \phi_{13}(z) \phi_{13}(1) \phi_{13}(\infty)\rangle  &= \langle  V_{\alpha_{13}}(0)V_{\alpha_{13}}(z)V_{\alpha_{13}}(1) V_{2\alpha_0-\alpha_{13}}(\infty)  Q_- Q_-  \rangle\nonumber\\ 
  &= z^{2 \alpha_{13}^2}(1-z)^{2 \alpha_{13}^2} J(z)  \label{4ep}
  \eea
  where 
  \be
 J(z) =\oint_{C_1}dt_1\oint_{C_2} dt_2 \, \frac{(t_1-t_2)^{2\alpha_{13}^2}}{  [t_1(t_1-1)(t_1-z)]^{2\alpha_{13}^{2}}   [t_2(t_2-1)(t_2-z)]^{2\alpha_{13}^{2}}    }    \label{4pt13}
 \ee
where we used $\alpha_{-}=-\alpha_{13}$, and note that   $\alpha_{13}^2=p/(1+p)$. 
In \reef{4ep} we have used conformal symmetry to set three of the points to $\{0,1,\infty\}$.
In \reef{4pt13} we need to specify the  closed contours $C_1\times C_2$ encircling the branch-points at $t_1,t_2 = 0,1,z,\infty$. 
There are three independent contours, and therefore \reef{4ep} is given by a particular linear combination of these three closed contours. 
The three independent contours, are in one-to-one correspondence with the three solutions of the third order differential equation satisfied by $J(z)$:
\bea
z^2(z-1)^2&  J^{\prime\prime \prime}(z)  + q_2(z)\,  J^{\prime\prime}(z)  +   q_1(z) \,  J^{\prime}(z) +  q_0(z)\, J(z) =0  \label{diff1}
\eea
where $q_2(z)= 10 \alpha_{13}^2 (z-1) z (2 z-1)  $, $q_1(z)=2 \alpha_{13}^2 [12 \alpha_{13}^2+(62 \alpha_{13}^2-7) (z-1) z-2 ] $,
and $q_0(z)=4 \alpha_{13}^2 (5 \alpha_{13}^2-1) (6 \alpha_{13}^2-1) (2 z-1) $.
The three solutions of \reef{diff1} are identified with $J_i$ in \reef{tri4pt}.
The final step is to determine the coefficients $\alpha_i$ in \reef{tri4pt}, 
which is done as follows.   The correlation function  $\langle \phi_{13}(0) \cdots \phi_{13}(z,\bar z) \phi_{13}(1)\phi_{13}(\infty) \rangle$    is single valued, which implies that there are no non-trivial monodromies as $z$ encircles $0$ and $1$. Requiring the absence of such monodromies determines the linear combination $\sum_{i=1}^3 \alpha_i |J_i(z)|^2$ of the three solutions of \reef{diff1} (up to an overall normalisation factor), and corresponds to selecting integration contours in \reef{4pt13} and in their anti-holomorphic $\langle \phi_{13}(\bar z)\dots \rangle$ counterparts. Finally, the normalisation factor can be fixed by taking the $z\rightarrow0$ of the correlator and comparing with the OPE.

\section{Perturbation Theory with Hamiltonian Truncation regulator}
\label{CPTHT}

Next we explain in detail how the third and fourth order perturbation theory corrections to the ground state energy are calculated when Hamiltonian Truncation is used instead as the regulator. This computation will serve as an example of the breakdown of locality that we argued for in section \ref{localHT}.

\subsection{Third order correction}

We start by computing \reef{simeq1}, the coefficient for the Tricritical Ising model ${\cal M}_{p=4}$ using Hamiltonian Truncation. This coefficient can be expressed as the product of matrices shown in \reef{pert1}, however it is more convenient to calculate it starting from the integral \reef{c3}.
 To do this, we radially order the integral of the three-point function in \reef{twothree}:
\be
3! \,  C \int\displaylimits_{0\leq |z_1|\leq |z_2|\leq 1}  \frac{d^2z_1}{|z_1|^{2-\Delta}}  \frac{d^2z_2}{|z_2|^{2-\Delta}}  \, \frac{1}{(z_1-z_2)^\frac{\Delta}{2} (1-z_2)^\frac{\Delta}{2} (1-z_1)^\frac{\Delta}{2}}  \times \text{c.c}.
\ee
where c.c. means multiplying the function by its complex conjugate.
Next we  expand the integral in the variables $z_1/z_2$ and $z_2$, using the series expansion 
 $
(1-\eps )^{-\Delta/2}=\sum_{k=0}^\infty r_k^{\Delta/2}\, \eps^k  
$, where 
\be
r_k^P \equiv \frac{ \Gamma(P+k)}{\Gamma(P)k!}  \, . \label{rdef}
\ee 
Integrating each term in the series leads to the following result
\be
3! (2\pi)^2 \,  C  \sum_{K=0}^\infty\sum_{Q=0}^\infty  \, \frac{ (A_{K,Q})^2}{(\Delta +2K)(\Delta+2Q)}  \label{HT3}
\ee
where the coefficients are given by the finite sums 
$
A_{K,Q}=  \sum_{k =0  }^{\text{min}\{K,Q\}} r^{\Delta/2}_{k} r^{\Delta/2}_{Q-k} r^{\Delta/2}_{K-k}
$.
Then, a Hamiltonian Truncation regularisation of \reef{HT3} proceeds by truncating the sums so that neither of the two denominator factors may exceed $\D$. We compare the Hamiltonian Truncation result in the $\D\rightarrow\infty$ limit with \reef{simeq1}. Taking this limit as a numerical extrapolation we find convergence to the analytical result. The convergence is slow because finite $\D$ corrections decouple as $\D^{3\Delta-2d}=\D^{-2/5}$ at large values of the cutoff.

\subsection{Fourth order corrections}

\label{calc1}

Next we compute the fourth order corrections. 
Again, these corrections can be expressed as products of matrices as shown in \reef{pert1}, but we find it more convenient to do the calculation starting from the integrals \reef{simeq2} and \reef{simeq3}.
For the Tricritical Ising case ${\cal M}_{p=4}$, the coefficient $\alpha_1$ in \reef{tri4pt} is zero, and thus we  have
\be
\langle \phi_{13}(x_1) \phi_{13}(x_2)\phi_{13}(x_3) \phi_{13}(1)\rangle_c = \alpha_2 |t_2(z_i)|+ \alpha_3 |t_3(z_i)|  - \text{disconnected} \label{schem1}
\ee
where the disconnected terms are written explicitly in \reef{c4}.
Next we consider the radially ordered  (r.o.) integral of each term in \reef{schem1}
\be
4!  \int_\text{r.o.} \prod_{i=1}^3\frac{d^2z_i }{|z_i|^{2-\Delta}  } |t_m(z_i)|  =  4! \int_\text{r.o.}   \prod_{i=1}^3\frac{d^2z_i }{|z_i|^{4/5}  }
\,  \frac{ z_{13}^{-22/5}(z_2-1)^{-22/5}}{[z_{12}z_{23}(z_3-1)(1-z_1)]^{-8/5}}\,  J_m(\eta) \times \text{c.c}. \label{pain1}
\ee
for $m=2,3$.
The functions $J_m$ in \reef{p4I} admit the following representation
$
J_2(\eta) = \eta^{-11/5}\sum_{j=0}^\infty  \hat{b}_j \eta^j
$, and $
J_3(\eta) = \eta^{-14/5}\sum_{j=0}^\infty  \hat{c}_j \eta^j
$.
The $\hat{b}_j$'s and $\hat{c}_j$'s can be obtained by either taking the series expansion of \reef{p4I}, or alternatively by deriving a recursion relation for the coefficients by inserting these series into the ODE  discussed in \reef{CG}. See appendix~\ref{afourpt} for further details. 

Next we perform a series expansion in the quantities $z_{i}/z_{i+1}$ and $z_3$ for all the terms in \reef{pain1}. 
This type of expansion was used to analyse perturbations to the $d=2$ critical Ising model in \cite{saleur1987two}.
After performing such an expansion for $|t_2|$ we are left with the following radially ordered integral 
\be
4!  \sum_{j,\bar j, k_i,\bar k_i=0}^\infty   \hat{b}_j \,   \hat{b}_{\bar j} \,     {\cal R} \,  \bar{\cal R}    
  \int_\text{r.o.}  \prod_{i=1}^3\frac{d^2z_i }{|z_i|^{4/5}  }
  \left( \frac{z_1}{z_2}\right)^{k_1+k_3+k_5}  
    \left(\frac{z_2}{z_3} \right)^{k_1+k_2+k_5 +k_6+j-\frac{3}{5}} 
    z_3^{k_1+k_4+k_6 -\frac{3}{5}}   
      \times  \text{c.c}.  \,  , \,  \label{toint1}
\ee
where 
 $
 {\cal R}=\prod_{i=1}^2 r_{k_i}^{-\frac{8}{5}}
 r_{k_{i+2}}^{\frac{3}{5}-j}
 r_{k_{i+4}}^{\frac{11}{5}+j} \,
 $,
and $\bar{\cal R}$ is obtained by replacing $\{k_i,j\}\rightarrow \{\bar k_{ i},\bar j\}$ on the indices in ${\cal R}$. Recall that the quantity $r_{k}^P$ is defined in \reef{rdef}.~\footnote{ We are left with an integral of the form
 $\int_\text{r.o.} d^2z_i \, z_1^a z_2^b z_3^c \times \text{c.c}.$ The  integrals over $\int_0^{|z_2|}d^2z_1$, $\int_0^{|z_3|}d^2z_2$ and  $\int_0^1 d^2z_3$ are  zero unless $c=\bar c$,  $a+b=\bar a + \bar b$ and  $a+b+c=\bar a + \bar b + \bar c$, respectively.   }
After performing few elementary integrals we are led to
\be
4! (2\pi)^3 \sum_{P,Q,R=0}^\infty  \frac{(B_{P,Q,R})^2}{(6/5+2P)(6/5+2Q)(6/5+2R)} \, .   \label{t2}
 \ee
 where  
 $
 B_{P,Q,R}=\sum_{q_i,j=0} \hat{b}_j r_{q_1}^{-\frac{8}{5}} r_{Q-q_1-q_2-q_3-j}^{-\frac{8}{5}}\, r_{P-q_1-q_2}^{\frac{3}{5}-j}r_{R-q_1-q_3}^{\frac{3}{5}-j}\, r_{q_2}^{\frac{11}{5}+j}r_{q_3}^{\frac{11}{5}+j} 
 $, and the lower indices on the $r$'s are required to be positive, which ensures that for any value of $\{P,Q,R\}$ there are a finite number of terms in the sum. This representation of the $t_2$ integral in \reef{pain1} is very convenient for implementing the Hamiltonian Truncation cutoff. 
 The denominators  $(6/5+2n)$, $n\in\mathds{N}$, in \reef{t2}  are interpreted as the energy of the states in the channel $[\phi_{13}]\times [\phi_{13}]\sim[\phi_{13}]$.
 The channel $[\phi_{13}][\phi_{13}]\sim[\mathds{1}]$ is accounted for in the $t_3$ term as we show next. 
  
The integral \reef{pain1} over $|t_3|$ proceeds through analogous steps. 
We expand in the quantities $z_{i}/z_{i+1}$ and $z_3$, which have magnitude less than 1 in the radially ordered integral, and then integrate. At this point we  notice an important difference with respect to  the previous calculation. The radially ordered integral of $t_3$ formally gives
\be
4! (2\pi)^3 \sum_{P,Q,R=0}^\infty  \frac{(C_{P,Q,R})^2}{(6/5+2P)( 2Q+\eps)(6/5+2R)} \, .   \label{t3}
 \ee
with $C_{P,Q,R}= \sum_{q_i,j} \hat{c}_jr_{q_1}^{-\frac{8}{5}} r_{Q-q_1-q_2-q_3-j}^{-\frac{8}{5}}\, r_{P-q_1-q_2}^{\frac{6}{5}-j}r_{R-q_1-q_3}^{\frac{6}{5}-j}\, r_{q_2}^{\frac{8}{5}+j}r_{q_3}^{\frac{8}{5}+j}$, where the lower indices  on the $r$'s are again required to be positive.
Equation \reef{t2} possesses an IR divergence due to the presence of a zero mode at $Q=0$, which we regulate with  $\eps$. 
The integral of \reef{schem1}, is finite and therefore we expect this singularity to be canceled by the disconnected terms.

Next we must evaluate the disconnected pieces. Two of them were computed already in section~\ref{HT1}, and we only need to compute the integral of $\langle \phi_{13}(x_1) \phi_{13}(x_2)\rangle $ $  \langle\phi_{13}(x_3)\phi_{13}(1)\rangle$ here.  
Radially ordering the integral and performing the series expansion of the integrand for $|z_i|\ll |z_{i+1}|\ll 1$ yields.
\be
 4!\sum_{k_i,\bar k_i=0}^\infty  \prod_{j=1}^2 r_{k_j}^{6/5} r_{\bar k_j}^{6/5}    \int_\text{r.o.}  \frac{d^2z_1}{|z_1|^{4/5}} 
    \frac{d^2z_2}{|z_2|^{4/5}}  \frac{d^2z_3}{|z_3|^{4/5}} 
\,  \left(\frac{z_1}{z_2}\right)^{k_1}
z_2^{-\frac{6}{5} +\eps}
\,   z_3^{k_2} 
\times c.c. \,   \label{ex22}
\ee
where we have introduced a small infrared regulator $\eps$. The integral simplifies to 
\be
\frac{1}{\eps} \times 4! (2\pi)^3 \sum_{k_1,k_2} \frac{(r_{k_1}^{6/5})^2}{6/5+2k_1}  \, \frac{(r_{k_2}^{6/5})^2}{6/5+2k_2}
\ee
which cancels against the zero mode $Q=0$ in \reef{t2}, because $\alpha_3 \, \hat{c}_0^2=1$. This is expected from the limiting behaviour of the four-point function in the  $x_1\rightarrow x_2$ limit.

After collecting the various terms, we are left with the following expression for the fourth order correction to the ground state energy in the Tricritical Ising model
\bea
\frac{c_4}{4!(2\pi)^3} =&- \sum_{P,Q,R=0}^\infty  \alpha_2 \frac{(B_{P,Q,R})^2}{(\frac{6}{5}+2P)\, (\frac{6}{5}+2Q)\, (\frac{6}{5}+2R)}-\sum_{\substack{P,R=0\\ Q=1}}^\infty  \alpha_3 \frac{(C_{P,Q,R})^2}{(\frac{6}{5}+2P)\, 2Q\, (\frac{6}{5}+2R)}  \nonumber \\
&+   \sum_{P,Q=0}^\infty \frac{1}{\frac{6}{5}+2P} \, \frac{(D_{P,Q})^2}{\frac{12}{5} +2P+2Q} \, \frac{1}{\frac{6}{5}+2Q}
+   \sum_{P,Q=0}^\infty  \frac{(D_{P,Q})^2}{\left(\frac{6}{5} +2P\right)^2 \left(\frac{12}{5}+2P+2Q\right)}  \, , 
\label{p44}
\eea
where $B_{P,Q,R}$ and $C_{P,Q,R}$ are defined right after \reef{t2} and \reef{t3}, respectively, and $D_{P,Q}=r_{P}^{6/5}r_{Q}^{6/5}$. The last two sums  in \reef{p44} arise from the integrals of the disconnected terms $\langle \phi_{13}(z_1) \phi_{13}(z_3)\rangle$ $\langle \phi_{13}(z_2) \phi_{13}(1)\rangle$ and
$\langle \phi_{13}(z_2) \phi_{13}(z_3)\rangle\langle \phi_{13}(z_1) \phi_{13}(1)\rangle$, respectively. At this point it is tempting to combine the terms in \reef{p44} right away into $\sum_{P,Q}  D_{P,Q}^2 \left(\frac{6}{5} +2P\right)^{-2} \left(\frac{6}{5}+2Q\right)^{-1}$, and identify them with the subtraction terms in \reef{pert4}.

While  $c_4$ is a finite quantity in this theory, each individual sum in 
\reef{p44} is divergent because the disconnected correlators diverge for $\Delta > d/2$. 
In general caution is required when performing  formal manipulations that add and subtract these sums. 
We regulate the sums using the Hamiltonian Truncation regulator, which simply consists in cutting off each  sum independently in \reef{p44}.

Next we quote the result for the $\phi_{13}$ perturbation of the $p=5$ diagonal minimal model. This computation proceeds through the same steps: in order to get an expression that we can regulate with the Hamiltonian Truncation cutoff, we radially order the integral, and perform all the series expansions and then integrate. 
Here we quote the final result
\bea
\frac{c_4}{4!(2\pi)^3} =&- \alpha_1\sum_{P,Q,R=0}^\infty  \frac{1}{ \frac{4}{3}+2P } \frac{(A_{P,Q,R})^2}{6+2Q}  \frac{1}{\frac{4}{3}+2R  } 
-\alpha_2\sum_{P,Q,R=0}^\infty  \frac{1}{ \frac{4}{3}+2P }\frac{(B_{P,Q,R})^2}{\frac{4}{3}+2Q} \frac{1}{ \frac{4}{3}+2R } \nonumber \\ 
&-\alpha_3\sum_{\substack{P,R=0\\ Q=1}}^\infty   \frac{1}{ \frac{4}{3}+2P }   \frac{(C_{P,Q,R})^2}{2Q}  \frac{1}{ \frac{4}{3}+2R }  \nonumber \\%
&+   \sum_{P,Q=0}^\infty \frac{1}{\frac{4}{3}+2P} \, \frac{(D_{PQ})^2}{\frac{8}{3} +2P+2Q} \, \frac{1}{\frac{4}{3}+2Q}
+   \sum_{P,Q=0}^\infty  \frac{(D_{PQ})^2}{\left(\frac{4}{3} +2P\right)^2 \left(\frac{8}{3}+2P+2Q\right)}   \,  . 
\label{p54}
\eea
The coefficients $A$, $B$ and $C$ are given in the appendix~\ref{atensors}, while now $D_{PQ}=r_{P}^{4/3}r_Q^{4/3}$. To simplify the notation, we have carried out the $p=4$ and $p=5$ computations individually, even though the parallelisms are evident because we are perturbing by  $\phi_{13}$ each model. One difference between \reef{p54}  and  \reef{p44} is that $[\phi_{13}]\times [\phi_{13}]=[\mathds{1}]+[\phi_{13}]+[\phi_{15}]$ has three blocks in its own OPE for $p=5$. This reflects in a non-zero contribution for the $\alpha_1 \neq 0$ piece that corresponds to the exchange of $\phi_{15}$, which has  dimension 6. 

\subsection{Discussion}
\label{disc1}

Starting with the Tricritial Ising Model ${\cal M}_{p=4}$, we regulate the fourth order correction using Hamiltonian Truncation by comparing \reef{p44} with the fourth order term of the RS series given in \reef{pert4}. 
Focusing on the first three terms of \reef{p44} we note that each of them has three denominator factors. They should be identified with the $\Delta_{k}$ denominator factors  appearing in the first term of the RS expression of \reef{pert4}.
Therefore we identify the first three terms of \reef{p44} with the first term in \reef{pert4}. 
This identification implies that regulating using HT is a matter of truncating the sums of \reef{p44} so that each denominator factor is less than $\D$. 
Continuing with the   last two terms of \reef{p44},  
 we  combine  them and  identify it with the subtraction terms in \reef{pert4}.
All in all, we get the following Hamiltonian Truncation regulated expression
\bea
  \left.  \frac{c_4}{4!(2\pi)^3}  \right|_\text{HT}  =&- \sum_{P_i=0}^{\frac{6}{5}+2P_i \leq \D}  \alpha_2 \frac{(B_{P_1,P_2,P_3})^2}{(\frac{6}{5}+2P_1)\, (\frac{6}{5}+2P_2) (\frac{6}{5}+2P_3)} \nonumber\\
&-\sum_{\substack{P_i=0\\ Q=1}}^{\substack{  \frac{6}{5}+2P_i \leq \D  \\  \ 2Q \leq \D }}  \alpha_3 \frac{(C_{P_1,Q,P_2})^2}{(\frac{6}{5}+2P_1)\, 2Q\, (\frac{6}{5}+2P_2)}  \nonumber \\
&+ \sum_{P=0}^{\frac{6}{5}+2P \leq \D}  \frac{(r_{P}^{6/5})^2}{(\frac{6}{5}+2P)^2}    \sum_{Q=0}^{\frac{6}{5}+2Q \leq \D}  \frac{(r_{Q}^{6/5})^2}{\frac{6}{5} +2Q}\, .
\label{HT4}
\eea
We confirmed this identification of the various terms of  \reef{HT4} with \reef{pert4} by direct computation of the matrix elements $V$, and performing the matrix product multiplication. 

Next we would like to introduce an alternative truncation  (a.t.) of the sums in $c_4$,
\bea
  \left.  \frac{c_4}{4!(2\pi)^3}  \right|_\text{a.t.}  =  \left.  \frac{c_4}{4!(2\pi)^3}  \right|_\text{HT}  + 
 \underbrace{
  \Big( \sum_{P,Q=0}^{\frac{12}{5}+2P+2Q \leq \D} -
   \sum_{P,Q=0}^{\substack{  \frac{6}{5}+2P \leq \D \, ,   \\ \frac{6}{5}+2P \leq \D }} \Big)  \frac{(r_{P}^{6/5})^2}{(\frac{6}{5}+2P)^2}  \,    \frac{(r_{Q}^{6/5})^2}{\frac{6}{5} +2Q}  }_{(S_B-S_A)/ 4!(2\pi^3)}\, .
\label{HT5}
\eea
The a.t. differs from HT in the   subtraction terms:  instead of truncating these terms as in the last line of \reef{HT4}, the a.t. imposes the condition $\frac{12}{5}+2P+2Q \leq \D$ in the sums. 
We motivate this  truncation of the fourth order correction by our discussion of the divergences of the integrated four point function   in section \ref{HT1}.
There we argued that UV divergences  from the first term in \reef{pert4} 
behave as    $\sum_{P,Q=0}^{\frac{12}{5}+2P+2Q \leq \D}   \frac{(r_{P}^{6/5})^2}{(\frac{6}{5}+2P)^2}  \,    \frac{(r_{Q}^{6/5})^2}{\frac{6}{5} +2Q}$. 
Then,  the a.t.  ensures that  this divergence from the integrated four-point is cancelled by modifying  the subtraction term.

\begin{figure}[t]\centering
\includegraphics[width=7.25cm]{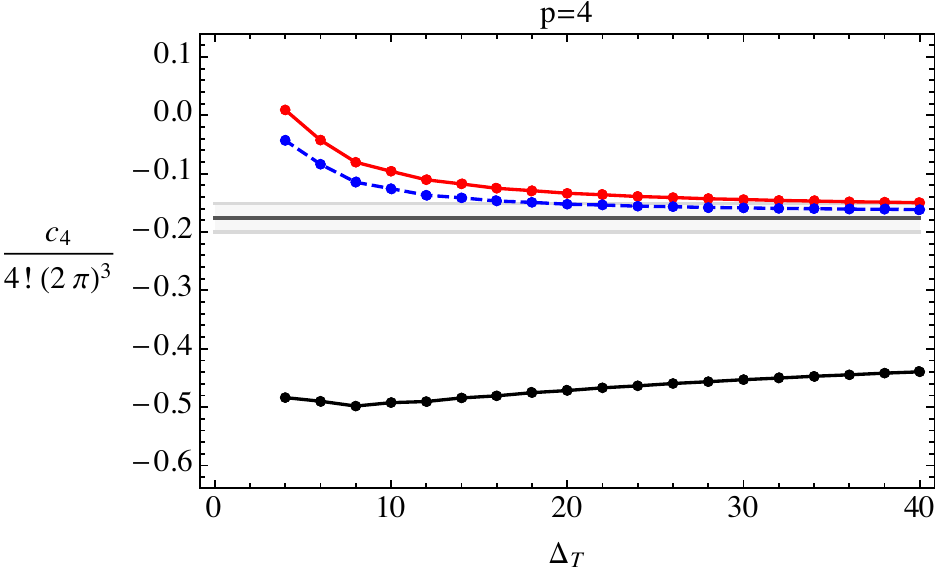} \ \ \ \ 
\includegraphics[width=7cm]{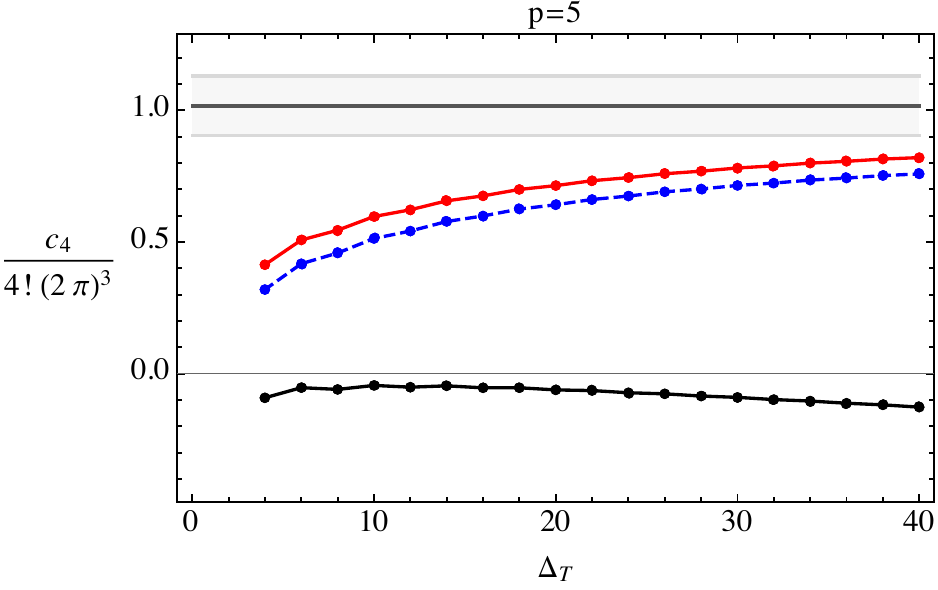}
\caption{Plotted in black and  red  are the results for the  HT  and alternative truncation (a.t.) regulation of the  $c_4$ coefficients respectively. The numerical integral of the connected four-point function is shown in gray. Plotted in blue are the HT regulated calculations plus the expressions in \reef{ppp1} (left plot) and \reef{ppp2} (right). \label{conv}  } 
\end{figure}

In the left panel of figure~\ref{conv} we plot  $c_4 / [4! (2 \pi )^3]$ as a function of $\D$ for the HT regulated expression  \reef{HT4} in black,  and for the  alternative regulator expression  \reef{HT5} in red. 
We also show in gray  the result of numerical  integration of the position space correlator in \reef{simeq2}.
The red line nicely converges to the value of the numerical integral. 
Instead, the  HT computation in black rises slowly towards the gray value, consistent with slow convergence. 
The  difference between HT  \reef{HT4} and  the alternative regulator \reef{HT5}  was calculated in section \ref{HT1} along with its asymptotic behaviour for large $\D$  in \reef{sab}. 
In the case of $p=4$, the difference between these two sums vanishes because 
$\Delta =6/5$ is lower than the threshold in \reef{redline} for $d=2$.
The blue line is obtained by taking the HT result  \reef{HT4} and adding to it 
\be
\frac{5  \, \Gamma \left(\frac{2}{5}\right)^2/\Gamma \left(\frac{4}{5}\right)- 25/2}{12\ 2^{4/5} \Gamma \left(\frac{6}{5}\right)^4 }   \, \D^{-1/5}
\approx  0.58 \, \D^{-1/5}\, ,  \label{ppp1}
\ee
which is obtained from \reef{sab}.  
This additional term vanishes in the $\D\rightarrow \infty$ limit, and therefore the blue and black line converge to the same point. 
The blue line tracks  the red line closely and seems to converge to the numerical value in gray.~\footnote{In figure~\ref{conv} we only plot even values for the truncation level. We find oscillations between even and odd truncation levels, which can be  ameliorated by a judicious choice of the cutoff, e.g. $  \sum^{2 \Delta +2P+2Q \leq \D}  \longrightarrow  \sum^{3 \Delta +2P+2Q \leq \D} $ in  \reef{HT5}. }
Therefore we understand the slow convergence of the black line  as a consequence of the small power for $\D$ of  $-1/5= 4(\Delta -d/2-1)$, for $d=2$ and $\Delta=6/5$, in the    $\D\rightarrow \infty$ limit.

Next we repeat the same exercise for the $\phi_{13}$ perturbation of the $p=5$ minimal model. The essential points are the same, and the full expression is relegated to appendix~\reef{adet}. 
The first three terms of \reef{p54} are given by the same expression regardless of whether we use the HT or the a.t. regulator. 
The only difference between regulators arises in the last two terms of \reef{p54}.
The difference between HT and a.t. diverges as $\D\rightarrow \infty$ because  
$\Delta =4/3$ is above the threshold in \reef{redline} for $d=2$.
This effect is shown in the right panel of figure~\ref{conv}. 
There we plot in black and red the HT and the a.t.  calculation, respectively.
The blue line is obtained by adding to the black line HT computation the following piece 
\be
\frac{6 \, \Gamma \left(\frac{2}{3}\right)^2-3\, \Gamma \left(\frac{1}{3}\right) }{16\;2^{1/3} \Gamma \left(\frac{4}{3}\right)^5}  \,  \D^{1/3} \approx 0.26 \, \D^{1/3}\, , \label{ppp2}
\ee
which is obtained from \reef{sab}. 
The blue and the red line show a nice convergence towards the gray numerical integral. 
For this example, the difference between the a.t. (red) and HT (black) regulators does not vanish in the limit $\D\rightarrow \infty$.
This was anticipated in the general discussion of figure~\ref{dimsfig}.

In this section we have developed conformal perturbation theory for two minimal models 
that sit at either site of the bound 
 $
\Delta \geq  d/2+1/4
$.
We have calculated the ground state energy up to fourth order in the coupling and we have confirmed our expectation that convergence is lost when the HT regulator is used for perturbations above the bound. We attribute the loss of convergence to a non-cancellation between  divergences of the integral of the $n$-point function 
and the divergences of its disconnected   pieces. This cancellation is guaranteed when using a  local regulator, but it can be lost when a non-local regulator, such as HT, is employed. Because this cancellation also appears in conformal perturbation theory for the energy gaps, the energy gaps will also be affected.

An earlier computation of the spectrum of the Tricritial Ising perturbed by $\phi_{13}$ using HT was done in \cite{Giokas:2011ix}. 
There the authors obtained precise results for the  energy levels on the cylinder. 
For the Tricritical Ising Model,  we have not found any difference between the  local and HT regulators in the $\D\rightarrow \infty$ limit.
Therefore the results of   \cite{Giokas:2011ix} are consistent with our analysis. 

\section{Conclusions}
\label{conc}

In this work, we further develop the theory of Hamiltonian Truncation (HT) for relevant  perturbations that are UV divergent ($d/2\le\Delta< d$). Since HT acts as an unconventional, non-local UV regulator, our aim is to clarify the conditions necessary for ensuring that the HT procedure yields UV finite results. To that end, we analyse conformal perturbation theory on the cylinder up to fourth order in the coupling.

We have shown that the finiteness of the integrated connected correlation functions,  appearing in conformal perturbation theory can depend on whether a local or HT regulator is used. The reason for this is the presence of delicate cancellations between singularities of the correlation function and singularities of its disconnected parts. 
When these connected correlators are integrated using a local regulator, such cancellations are preserved leaving a finite integral, but this fails when a non-local regulator such as HT is used. We have demonstrated this effect with explicit calculations of the ground state energy, but excited state energies will be similarly affected because they are also expressible in terms of integrals over connected correlation functions.

This leads us to introduce a \emph{four-point test} of HT consistency. It is applied by computing the fourth order perturbative correction to the energy of a state $E^{(4)}_i$ using HT and a local regularisation and constructing the following quantity
\be
\delta_i(\D)\equiv  [E^{(4)}_i]_\text{HT}   -  [E^{(4)}_i]_\text{local}  \, .   \label{test}
\ee
In a consistent Hamiltonian Truncation formulation this difference must vanish when the $\D$  cutoff is sent to infinity. 
We showed that   the finiteness of $ \delta_i(\D)$ critically depends on the dimension $\Delta$ of the perturbing operator. When $ \delta_i(\D)$ diverges  as $\D\rightarrow \infty$, this signals a  breakdown of locality. This happens first at the fourth order in the  Rayleigh--Schr$\ddot{\text{o}}$dinger series. In this study, we evaluate \reef{test} in the ground state, and find $\delta_\text{g.s.}(\D)\sim \D^{4\Delta-2d-1}$. The exact coefficient of the four-point test $\delta_\text{g.s.}(\D)$ asymptotic behaviour at large $\D$ can be found in \reef{sab}.
 In particular, this calculation implies the need for non-local counter-terms in Hamiltonian Truncation.
After incorporating such counter-terms in HT the four-point test should  be satisfied.  
Counter-terms of this type have been worked out  in the context of the $\phi^4$ theory in $d=3$~\cite{Elias-Miro:2020qwz,Anand:2020qnp}, and more recently a theory to compute  all the HT counter-terms was proposed in~\cite{Cohen:2021erm}. 
In light of our results,  it would be interesting to formulate a Hamiltonian Truncation theory including all the necessary UV divergent counter-terms for  perturbations  with an arbitrary dimension in the range $d/2 +1/4\leq \Delta < d$.

We provided an example showing the   mismatch in \reef{test} between HT and local regularisation. We calculate the ground state energy of two CFTs deformed by an operator with dimension on either side of the threshold $d/2+1/4$ implied by the four-point test. Specifically, we considered the diagonal minimal models ${\cal M}_p$, with $p=4$ and 5, deformed by the operator $\phi_{13}$ with dimension $\Delta_p=2-4/(p+1)$. Both models have a well defined fourth order correction to the ground state energy, when a local regulator is employed. Yet, when regulating with HT, the fourth order ground state energy of the $p=5$ case diverges in the $\D\rightarrow \infty$ limit.

Although this work has been primarily focused on conformal perturbation theory, it would be very worthwhile to apply HT non-perturbatively to analyse a greater variety of QFTs, especially those in higher dimensions. Our results suggest that Hamiltonian Truncation can be applied in the range $d/2 \leq \Delta < d/2+1/4$ with minimal modifications, namely by using only the second order counter-term computed in \reef{ct0}-\reef{p1}.
As more conformal data relating to higher dimensional CFTs become available, it will become possible to analyse more QFTs using HT. The conformal bootstrap programme \cite{Rattazzi:2008pe,Poland:2018epd} promises to deliver such data. For example,    precise CFT data is already known for the 3D Ising~\cite{El-Showk:2012cjh,El-Showk:2014dwa,Simmons-Duffin:2016wlq} and 3D $O(2)$ model~\cite{Chester:2019ifh,Liu:2020tpf}. The latter model features a single relevant operator with dimension in the range $d/2 \leq \Delta_s < d/2+1/4$, falling precisely within the window where the four-point test \reef{test} is satisfied.

Another worthwhile goal for research is to improve the efficiency of Hamiltonian Truncation. One strategy for achieving improvements is to derive an effective truncated Hamiltonian whose spectrum converges faster in the $\D\rightarrow \infty$ limit~\cite{Hogervorst:2014rta,Giokas:2011ix,Rychkov:2014eea,Rychkov:2015vap,Elias-Miro:2015bqk,Elias-Miro:2017xxf,Elias-Miro:2017tup,Cohen:2021erm}.
So far this approach has been applied only for perturbations with dimension $\Delta < d/2$ -- with the exception of \cite{Giokas:2011ix}. We did not develop the theory of effective truncated Hamiltonians for $\Delta \geq d/2$ here. Nevertheless, the four-point test provides a consistency check for candidate effective truncated Hamiltonians when  the perturbing operator has dimension $\Delta \geq d/2$.

\subsection*{Acknowledgements}

We are grateful to Ed Hardy, Matthijs Hogervorst, Ami Katz, Markus Luty and Slava Rychkov for useful discussions and comments on the draft.


\appendix

\section{Energy gaps}
\label{agaps}

In this appendix we give  details on the derivation of \reef{pert3}.
On general grounds we  have
\be
E_i - E_0 =  - \lim_{\tau \rightarrow \infty} \frac{1}{\tau} \log \left[ \langle  O_i(\tau/2)O_i(-\tau/2)\rangle_{g} - \langle O_i(0)\rangle_{g}^2  \right] \label{agap1}
\ee
and    $\langle O_i(x) \cdots \rangle_g \equiv  \langle  O_i(x) \cdots   e^{-g \int {\cal L}_\text{int}} \rangle/  \langle    e^{-g \int {\cal L}_\text{int}} \rangle$, where the correlations $\langle \cdots \rangle$ without subscript are  computed in the  unperturbed $g=0$ vacuum. 
Formula \reef{agap1} follows from the large $\tau$ limit of
\begin{align*}
 \langle  O_i(\tau/2)O_i(-\tau/2)\rangle_{g} & = \sum_n  \langle 0 |  O_i(\tau/2)|n\rangle \langle n|O_i(-\tau/2)|0\rangle_{g}\\ & = \langle O_i(0)\rangle_g^2+e^{-(E_i-E_{gs})\tau }|\langle 0|O_i | 1 \rangle_g |^2 +  \cdots
\end{align*}
where the dots denote exponentially suppressed terms.  Then, we get   \reef{agap1}
by  taking the logarithm  and the large $\tau$ limit of the last formula. 

Next we evaluate \reef{agap1} two second order in perturbation theory. We will do so for the perturbation ${\cal L}_\text{int}= \int_\text{cyl} \phi_\Delta$ of a CFT on a cylinder   $\lim_{\tau \rightarrow \infty}[-\tau/2,\tau/2]\times S_R^{d-1}$. 
Expanding $\log \left[ \langle O_i(\tau/2)O_i(-\tau/2)  \rangle_g - \langle O_i(0) \rangle_g^2 \right]$ to second order gives 
\bea
&\log   \langle O_i(\tau/2)O_i(-\tau/2)\rangle  -  g \int_\text{cyl} d^d \xi\frac{ \langle O_i(\tau/2)  \phi_\Delta (\xi)  \,  O_i(-\tau/2)\rangle}{   \langle O_i(\tau/2)   O_i(-\tau/2)\rangle  }\nonumber  \\[.2cm]
&+ \frac{g^2}{2}  \int_\text{cyl} d^d \xi_1 d^d \xi_2  \frac{ \langle O_i(\tau/2)  \phi_\Delta (\xi_1)  \phi_\Delta (\xi_2)  \,  O_i(-\tau/2)\rangle_c}{   \langle O_i(\tau/2)   O_i(-\tau/2)\rangle  } \nonumber\\ &-\frac{g^2}{2} \left\{    \int_\text{cyl} d^d \xi  \frac{ \langle O_i(\tau/2)  \phi_\Delta (\xi)  \,  O_i(-\tau/2)\rangle_c}{   \langle O_i(\tau/2)   O_i(-\tau/2)\rangle  } \right\}^2 \label{tolimit}
\eea
where as usual the connected correlator is given by
\bea
 \langle O_{\tau/2}  \phi_{\xi_1} \phi_{\xi_2}  \,  O_{-\tau/2}\rangle_c = \langle O_{\tau/2}  \phi_{\xi_1} \phi_{\xi_2}  \,  O_{-\tau/2}\rangle 
 & - \langle O_{\tau/2}  \phi_{\xi_1}  \rangle \, \langle \phi_{\xi_2}  \,  O_{-\tau/2}\rangle \nonumber\\
 & - \langle  \phi_{\xi_1}  \phi_{\xi_2}  \rangle \, \langle   O_{\tau/2}  \,  O_{-\tau/2}\rangle \nonumber\\
 & - \langle O_{\tau/2}  \phi_{\xi_2}  \rangle \, \langle \phi_{\xi_1}  \,  O_{-\tau/2}\rangle    \,  ,
\eea
with   the notation $\phi_\Delta(\xi)\equiv  \phi_{\xi}$ and $O_i(x)= O_x$.
Next we take the limit $\lim_{\tau\rightarrow \infty} \frac{-1}{\tau} \reef{tolimit}$. 

The first term is given by
\be
\lim_{\tau\rightarrow \infty}\frac{-1}{\tau} \log   \langle O_i(\tau/2)O_i(-\tau/2)\rangle  =\lim_{\tau\rightarrow \infty}\frac{2\Delta_i }{\tau} \log  |e^{\frac{\tau}{2 R}}-e^{-\frac{\tau}{2R}}| =  \Delta_i/R \,  ,
\ee
as expected in the unperturbed theory $E_i-E_0= \Delta_i/R+O(g)$.

To compute the second term of \reef{tolimit} we perform a Weyl transformation to  map the correlators of the cylinder into the plane, 
\be
 \int_\text{cyl} d^d \xi\frac{ \langle O_i(\tau/2)  \phi_\Delta (\xi)  \,  O_i(-\tau/2)\rangle}{   \langle O_i(\tau/2)   O_i(-\tau/2)\rangle  } =\int_{\mathds{R}^d} d^dx \left|\frac{x}{R}\right|^{\Delta-d} \frac{ \langle O_i(\tau/2)   \phi_\Delta (x)  \,  O_i(-\tau/2)\rangle}{   \langle O_i(\tau/2)   O_i(-\tau/2)\rangle  } \label{prev} \, . 
\ee
where the correlators on the l.h.s. are evaluated on the cylinder   while those on the right are evaluated on $\mathds{R}^d$. 
We take the large $\tau$ limit
\begin{align}
\lim_{\tau \rightarrow \infty}\frac{g}{\tau} \reef{prev}= \lim_{\tau \rightarrow \infty} \frac{g}{\tau} \int d^d x \left| \frac{x}{R} \right|^{\Delta-d}
  \frac{| e^{\frac{\tau}{2 R}}-e^{-\frac{\tau}{2R}}|^{2\Delta_i} \,\,C_{OO}^\phi}{|e^{\frac{\tau}{2R}}-x|^{\Delta}|x-e^{-\frac{\tau}{2R}}|^{\Delta}|e^{\frac{\tau}{2R}}-e^{-\frac{\tau}{2R}}|^{2\Delta_i-\Delta}} \nonumber
 \end{align}
and note that the region of the integrand $x\sim  e^\frac{\tau}{2R}$ does not contribute.~\footnote{We have used a simplified notation so that  $|x-e^{-\frac{\tau}{2R}}|$ means $|\vec x-e^{-\frac{\tau}{2R}}\vec n|$, where $\vec n$ is a unit $d$-dimensional vector. At large $\tau$, $|\vec x-e^{-\frac{\tau}{2R}}\vec n|\sim |\vec x|\equiv |x|$. }
 Therefore we are left with 
 \be
\lim_{\tau \rightarrow \infty} \frac{g}{\tau}  \int     d^d x \left| \frac{x}{R} \right|^{\Delta-d} 
  \frac{C_{OO}^\phi}{|x|^{\Delta}}   = C_{OO}^\phi \lim_{\tau \rightarrow \infty} \frac{g}{\tau} \frac{1}{R^{\Delta-d}} \int_{r=Re^{-\frac{\tau}{2R}}}^{r=Re^{+\frac{\tau}{2R}}}      \frac{dr}{r} \int dS_{d-1}   =C_{OO}^\phi \frac{gS_{d-1}}{R^{\Delta-d+1}}  \, . 
\ee

Finally, the third term of \reef{tolimit}. We use translation invariance  to shift the first field into the origin  $\phi_\Delta (\xi_1)\rightarrow \phi_\Delta (0)$, then the first integral trivializes $\int d^d\xi_1=$ \newline$\int_{-\tau/2}^{\tau/2}d\tau_1  \int  dS_{d-1} R^{d-1}= \tau R^{d-1}S_{d-1}$. Next we Weyl-map from the cylinder to the plane
\begin{multline}
 \int_\text{cyl} d^d \xi_2  \frac{ \langle O_i(\tau/2)  \phi_\Delta (0)  \phi_\Delta (\xi_2)  \,  O_i(-\tau/2)\rangle_c}{   \langle O_i(\tau/2)   O_i(-\tau/2)\rangle  }  \\=
 \int_{\mathds{R}^d} d^d x \left|\frac{x}{R}\right|^{\Delta-d}  \frac{ \langle O_i(\tau/2)  \phi_\Delta (R\vec{u})  \phi_\Delta (x)  \,  O_i(-\tau/2)\rangle_c}{  |e^{\frac{\tau}{2R}}-e^{-\frac{\tau}{2R}}|^{2\Delta_i}  }  \label{tolimit2}
\end{multline}
where $\vec{u}$ is a unit vector. Finally we take the limit
\be
\lim_{\tau \rightarrow \infty }-\frac{g^2}{2} R^{d-1}S_{d-1} \reef{tolimit2} =  -\frac{g^2}{2}  R^{d-1} S_{d-1} \int_{\mathds{R}^d} d^d x \left|\frac{x}{R}\right|^{\Delta-d}   \langle O_i(\infty)  \phi_\Delta (R\vec{n})  \phi_\Delta (x)  \,  O_i(0)\rangle_c  \, ,
\ee
where $O_i (\infty)\equiv  \lim_{x\rightarrow \infty}  x^{2\Delta_i }O_i(x)$.


\section{Four-point functions}
\label{afourpt}

\subsection{Constants}

For the minimal models ${\cal M}_p$ perturbed by the $\phi_{13}(x)$ operator, the general form of the four-point function for $\phi_{13}$ is shown in \reef{tri4pt}. In the case of the Tricritical Ising model ${\cal M}_4$, the constants $\{\alpha_i\}$ appearing in the four-point function take the values
\be
\alpha_1=0\, ,\qquad \alpha_2 = \Gamma\left(\tfrac{2}{5}\right)^2\, , \qquad \alpha_3 = \Gamma\left(\tfrac{2}{5}\right)^2.
\label{alphap4}
\ee
We also need the constants $\{u_i,v_i\}$ to specify the $J_i$ functions for the Tricritical Ising model, defined in \reef{p4I}. For the case of the $J_2$ function, the relevant constants are
\be
u_2=\frac{\sqrt{2+\sqrt{5}}}{2}\, ,\qquad v_2 = \frac{5^{1/4}}{\pi}.
\ee
For the case of the $J_3$ function, the required constants are
\be
u_3 = \frac{1-\sqrt{5}}{4}\, ,\qquad v_3 = \frac{5^{1/4}}{\pi}\sqrt{\frac{1+\sqrt{5}}{2}}\, .
\ee

Similarly, in the case of the $p=5$ minimal model, the constants $\{\alpha_i\}$ needed to specify the four-point function shown in \reef{tri4pt} take the values
\be
	\alpha_1=\frac{1}{36}\, ,\qquad
	\alpha_2=\frac{1}{18}\, ,\qquad
	\alpha_3=\frac{1}{36}\, .
\ee
{\vspace{2pt}}

\subsection{Recursion relations}
\label{arecursion}

In section \ref{cptex}, we derived a set of functions $J_i(\eta)$, which enter our results for four-point functions, as shown in Eq.~\reef{tri4pt}. The $J_i(\eta)$ functions can be expanded as a power series in $\eta$. We use the coefficients of these expansions in formulae such as \reef{HT4} for  calculating fourth order corrections to the ground state energy $c_4$ in our minimal model example theories. The coefficients can be found by directly expanding the $J_i(\eta)$ functions. However, finding coefficients of the expansion at very high orders using this method is inefficient. Instead, we find it useful to derive recursion relations, which enable more efficient computation of the higher order expansion coefficients.

For the case of the ${\cal M}_4$ Tricritical Ising model, the relevant $J_i(\eta)$ functions are shown explicitly in \reef{p4I}. We begin by considering the series expansion for $J_2(\eta)$. It has the following form
\be
J_2(\eta) = \eta^{-11/5}\sum_{n=0}^\infty\hat{b}_n\eta^n. 
\label{p4J2exp}
\ee
To derive a recursion relation for the coefficients $\hat{b}_n$, we can plug the expansion \reef{p4J2exp} directly into the ODE which determines $J_2$, shown in \reef{diff1}. Equating coefficients of $\eta^n$ yields the following result
\begin{multline}
	0=\left(760+2360(n-1)+1175(n-1)(n-2)+125(n-1)(n-2)(n-3)\right)\hat{b}_{n-1}\\
	+\left(456-600n-1350n(n-1)-250n(n-1)(n-2)\right)\hat{b}_n\\
	+\left(-240(n+1)+175n(n+1)+125n(n+1)(n-1)\right)\hat{b}_{n+1}.
\end{multline}
We apply this equation recursively to find each $\hat{b}_n$ for large $n$. To do this, we must first set $\hat{b}_{-1}=0$ and then input $\hat{b}_0=-2\pi/\left[5^{1/4}\Gamma(-2/5)\Gamma(1/5)\Gamma(8/5)\right]$, obtained by direct expansion of \reef{p4I}.

The four-point function of the Tricritical Ising model also contains a second function $J_3(\eta)$. It admits a similar series expansion:
\be
J_3(\eta) = \eta^{-14/5}\sum_{n=0}^\infty\hat{c}_n\eta^n.
\ee
Again, a recursion relation for the $\hat{c}_n$ can be derived by plugging the expansion above into the ODE in \reef{diff1}. We find
\begin{multline}
	0=\left(160+1310(n-1)+950(n-1)(n-2)+125(n-1)(n-2)(n-3)\right)\hat{c}_{n-1}\\
	+\left(144+300n-900n(n-1)-250n(n-1)(n-2)\right)\hat{c}_n\\
	+\left(-90(n+1)-50n(n+1)+125n(n+1)(n-1)\right)\hat{c}_{n+1}.
\end{multline}
To apply this formula, we first set $\hat{c}_{-1}=0$ and $\hat{c}_0=-1/\Gamma(2/5)$, obtained by direct expansion of \reef{p4I}. Then all subsequent $\hat{c}_n$ may be determined recursively.

Similar logic can be applied to the corresponding four-point function in the case of the ${\cal M}_5$ minimal model. For this model, the four-point function is made up of three new $J_i$ functions shown in \reef{p54pt}. They admit the series representations below
\bea
J_1(\eta)&=\sum_{j=0}^\infty\hat{a}_j \eta^j\, ,\label{J1}\\
J_2(\eta)&=\eta^{-7/3}\sum_{j=0}^\infty \hat{b}_j \eta^j\, ,\\
J_3(\eta)&=\eta^{-3}\sum_{j=0}^\infty \hat{c}_j \eta^j\, \label{J3}.
\eea
To derive recursion relations for the hatted coefficients, we input the expansions above into the defining ODE in \reef{diff1} as before. Starting with $J_1$, we find
\begin{multline}
	0=\left(760+670(n-1)+150(n-1)(n-2)+9(n-1)(n-2)(n-3)\right)\hat{a}_{n-1}\\+\left(-380-670n-225n(n-1)-18n(n-1)(n-2)\right)\hat{a}_{n}\\
	+\left(120(n+1)+75n(n+1)+9n(n+1)(n-1)\right)\hat{a}_{n+1}
\end{multline}
This can be used to determine the subsequent $\hat{a}_n$ recursively by first inputting $\hat{a}_{-1}=0$ and $\hat{a}_0=52/27$. 

For the $J_2$ case, we find
\begin{multline}
	0=\left(60+180(n-1)+87(n-1)(n-2)+9(n-1)(n-2)(n-3)\right)\hat{b}_{n-1}\\
	+\left(40-40n-99n(n-1)-18n(n-1)(n-2)\right)\hat{b}_n\\
	+\left(-20(n+1)+12n(n+1)+9n(n+1)(n-1)\right)\hat{b}_{n+1}.
\end{multline}
Again, we determine the subsequent $\hat{b}]_n$ by first inputting $\hat{b}_{-1}=0$ and $\hat{b}_0=4$.

For the $J_3$ case, we obtain
\begin{multline}
	0=\left(10+94(n-1)+69(n-1)(n-2)+9(n-1)(n-2)(n-3)\right)\hat{c}_{n-1}\\
	+\left(10+32n-63n(n-1)-18n(n-1)(n-2)\right)\hat{c}_n\\
	+\left(-6(n+1)-6n(n+1)+9n(n+1)(n-1)\right)\hat{c}_{n+1}.
	\label{chatp5}
\end{multline}
To determine the subsequent $\hat{c}_n$, we first input $\hat{c}_{-1}=0$ and $\hat{c}_0=-6$. A new subtlety is encountered when performing this calculation, because for $n=2$ \reef{chatp5} provides no constraint on $\hat{c}_3$. The inability of this procedure to fix $\hat{c}_3$ should be expected because of the similarity between the series expansions for $J_1$ in \reef{J1} and $J_3$ in \reef{J3}. Specifically it is possible to take linear combinations of $J_1$ and $J_3$ type series solutions to the ODE in \reef{diff1} to form new independent solutions of the $J_3$ type. We eliminate this ambiguity by also inputting the value $\hat{c}_3=-920/27$. This value may be derived by directly expanding $J_3$ in \reef{p54pt}. The remaining $\hat{c}_n$ coefficients may now be derived by applying the recursion relation.

\subsection{Hamiltonian Truncation coefficients}
\label{atensors}

In this appendix, we give the $A_{P,Q,R}$, $B_{P,Q,R}$, and $C_{P,Q,R}$  coefficients of \reef{p54}, which characterise the fourth order correction to the ground state energy in the deformed ${\cal M}_5$ minimal model:
\bea
A_{P,Q,R}&=\sum_{q_i,j=0} \hat{a}_j r_{q_1}^{-\frac{5}{3}} r_{Q-q_1-q_2-q_3-j}^{-\frac{5}{3}}\, r_{P-q_1-q_2}^{-\frac{5}{3}-j}r_{R-q_1-q_3}^{-\frac{5}{3}-j}\, r_{q_2}^{\frac{14}{3}+j}r_{q_3}^{\frac{14}{3}+j}\, ,   \\[.2cm]
B_{P,Q,R}&=\sum_{q_i,j=0} \hat{b}_j r_{q_1}^{-\frac{5}{3}} r_{Q-q_1-q_2-q_3-j}^{-\frac{5}{3}}\, r_{P-q_1-q_2}^{\frac{2}{3}-j}r_{R-q_1-q_3}^{\frac{2}{3}-j}\, r_{q_2}^{\frac{7}{3}+j}r_{q_3}^{\frac{7}{3}+j}\, , \\[.2cm]
C_{P,Q,R}&=\sum_{q_i,j=0} \hat{c}_j r_{q_1}^{-\frac{5}{3}} r_{Q-q_1-q_2-q_3-j}^{-\frac{5}{3}}\, r_{P-q_1-q_2}^{\frac{4}{3}-j}r_{R-q_1-q_3}^{\frac{4}{3}-j}\, r_{q_2}^{\frac{5}{3}+j}r_{q_3}^{\frac{5}{3}+j} \, .
\eea
The $r^\Delta_j$ factors are defined in \reef{rdef}. In the sums above, the ranges of the indices being summed over are restricted, so that the lower indices $j$ of the $r^\Delta_j$ factors cannot be negative. This ensures that all the above are finite sums. 

The $\hat{a}_j$, $\hat{b}_j$ and $\hat{c}_j$ are coefficients appearing in series expansions for the $J_i(\eta)$ functions shown in \reef{p54pt}. The four-point function of the ${\cal M}_5$ minimal model is constructed in \reef{tri4pt} using these functions. The coefficients above can be determined by directly expanding the functions shown in \reef{p54pt}, or alternatively by using recursion relations, as explained in appendix~\ref{arecursion}.

\subsection{Regulation of $c_4$ for the $\phi_{13}$ perturbation of  ${\cal M}_{p=5}$}
\label{adet}

In this appendix we provide the regularization of  \reef{p54}.  Using  a Hamiltonian Truncaiton cutoff  leads to 
\bea
 \left.  \frac{c_4}{4!(2\pi)^3}  \right|_\text{HT}    =&
 - \alpha_1\sum_{P_i,Q=0}^{\substack{\frac{4}{3}+2P_i\leq \D\\ 6+2Q \leq \D }}  \frac{1}{ \frac{4}{3}+2P_1 } \frac{(A_{P_1,Q,P_2})^2}{6+2Q}  \frac{1}{\frac{4}{3}+2P_2  }\nonumber\\ 
&-\alpha_2\sum_{P_i=0}^{\frac{4}{3}+2P_i \leq \D}  \frac{1}{ \frac{4}{3}+2P_1 }\frac{(B_{P_1,P_2,P_3})^2}{\frac{4}{3}+2P_2} \frac{1}{ \frac{4}{3}+2P_3 } \nonumber \\ 
&-\alpha_3\sum_{\substack{P_i=0\\ Q=1}}^{\substack{\frac{4}{3}+2P_i\leq \D\\ 2Q \leq \D }}    \frac{1}{ \frac{4}{3}+2P_1 }   \frac{(C_{P,Q,R})^2}{2Q}  \frac{1}{ \frac{4}{3}+2P_2 }  \nonumber \\
&+   \sum_{P=0}^{\frac{4}{3}+2P\leq \D}  \frac{(r_{P}^{4/3})^2}{(\frac{4}{3}+2P)^2}   \sum_{Q=0}^{\frac{4}{3}+2Q\leq \D} \frac{(r_{Q}^{4/3})^2}{\frac{4}{3}+2Q} \, .
\eea
In the main text we have also introduced an alternative truncation (a.t.) of the fourth order calculation given by  
\bea
 \left.  \frac{c_4}{4!(2\pi)^3}  \right|_\text{a.t.}   =  \left.  \frac{c_4}{4!(2\pi)^3}  \right|_\text{HT}&  -   \sum_{P=0}^{\frac{4}{3}+2P\leq \D}  \frac{(r_{P}^{4/3})^2}{(\frac{4}{3}+2P)^2}   \sum_{Q=0}^{\frac{4}{3}+2Q\leq \D} \frac{(r_{Q}^{4/3})^2}{\frac{4}{3}+2Q} \\
& +   \sum_{P=0}^{\frac{8}{3}+2P+2Q\leq \D}  \frac{(r_{P}^{4/3})^2}{(\frac{4}{3}+2P)^2} \,   \frac{(r_{Q}^{4/3})^2}{\frac{4}{3}+2Q} \, . 
\eea
The coefficients $A$, $B$ and $C$ are given in the appendix~\ref{atensors}, while  $D_{PQ}=r_{P}^{4/3}r_Q^{4/3}$ and $r_K^\Delta$ is defined in \reef{rdef}. 


\small

\bibliography{biblio}

\providecommand{\href}[2]{#2}\begingroup\raggedright\begin{thebibliography}{10}

\bibitem{Yurov:1989yu}
V.~P. Yurov and {\relax Al}.~B. Zamolodchikov, ``{Truncated Conformal Space
  Approach to Scaling Lee-Yang Model},''
\href{http://dx.doi.org/10.1142/S0217751X9000218X}{{\em Int.J.Mod.Phys.}
  {\bfseries A5} (1990) 3221--3246}.

\bibitem{Yurov:1990kv}
V.~P. Yurov and A.~B. Zamolodchikov, ``{Correlation functions of integrable 2-D
  models of relativistic field theory. Ising model},''
  \href{http://dx.doi.org/10.1142/S0217751X91001660}{{\em Int. J. Mod. Phys. A}
  {\bfseries 6} (1991) 3419--3440}.

\bibitem{Lassig:1990xy}
M.~Lassig, G.~Mussardo, and J.~L. Cardy, ``{The scaling region of the
  tricritical Ising model in two-dimensions},''
  \href{http://dx.doi.org/10.1016/0550-3213(91)90206-D}{{\em Nucl. Phys. B}
  {\bfseries 348} (1991) 591--618}.

\bibitem{Katz:2016hxp}
E.~Katz, Z.~U. Khandker, and M.~T. Walters, ``{A Conformal Truncation Framework
  for Infinite-Volume Dynamics},''
  \href{http://dx.doi.org/10.1007/JHEP07(2016)140}{{\em JHEP} {\bfseries 07}
  (2016) 140}, \href{http://arxiv.org/abs/1604.01766}{{\ttfamily
  arXiv:1604.01766 [hep-th]}}.

\bibitem{Katz:2013qua}
E.~Katz, G.~Marques~Tavares, and Y.~Xu, ``{Solving 2D QCD with an adjoint
  fermion analytically},''
  \href{http://dx.doi.org/10.1007/JHEP05(2014)143}{{\em JHEP} {\bfseries 05}
  (2014) 143}, \href{http://arxiv.org/abs/1308.4980}{{\ttfamily arXiv:1308.4980
  [hep-th]}}.

\bibitem{Katz:2014uoa}
E.~Katz, G.~Marques~Tavares, and Y.~Xu, ``{A solution of 2D QCD at Finite $N$
  using a conformal basis},'' \href{http://arxiv.org/abs/1405.6727}{{\ttfamily
  arXiv:1405.6727 [hep-th]}}.

\bibitem{Fitzpatrick:2018ttk}
A.~L. Fitzpatrick, J.~Kaplan, E.~Katz, L.~G. Vitale, and M.~T. Walters,
  ``{Lightcone effective Hamiltonians and RG flows},''
  \href{http://dx.doi.org/10.1007/JHEP08(2018)120}{{\em JHEP} {\bfseries 08}
  (2018) 120}, \href{http://arxiv.org/abs/1803.10793}{{\ttfamily
  arXiv:1803.10793 [hep-th]}}.

\bibitem{Anand:2021qnd}
N.~Anand, A.~L. Fitzpatrick, E.~Katz, and Y.~Xin, ``{Chiral Limit of 2d QCD
  Revisited with Lightcone Conformal Truncation},''
  \href{http://arxiv.org/abs/2111.00021}{{\ttfamily arXiv:2111.00021
  [hep-th]}}.

\bibitem{Hogervorst:2018otc}
M.~Hogervorst, ``{RG flows on $S^d$ and Hamiltonian truncation},''
  \href{http://arxiv.org/abs/1811.00528}{{\ttfamily arXiv:1811.00528
  [hep-th]}}.

\bibitem{Hogervorst:2021spa}
M.~Hogervorst, M.~Meineri, J.~Penedones, and K.~S. Vaziri, ``{Hamiltonian
  truncation in Anti-de Sitter spacetime},''
  \href{http://arxiv.org/abs/2104.10689}{{\ttfamily arXiv:2104.10689
  [hep-th]}}.

\bibitem{Chen:2021bmm}
H.~Chen, A.~L. Fitzpatrick, and D.~Karateev, ``{Form Factors and Spectral
  Densities from Lightcone Conformal Truncation},''
  \href{http://arxiv.org/abs/2107.10285}{{\ttfamily arXiv:2107.10285
  [hep-th]}}.

\bibitem{Chen:2021pgx}
H.~Chen, A.~L. Fitzpatrick, and D.~Karateev, ``{Bootstrapping 2d $\phi^4$
  Theory with Hamiltonian Truncation Data},''
  \href{http://arxiv.org/abs/2107.10286}{{\ttfamily arXiv:2107.10286
  [hep-th]}}.

\bibitem{Hogervorst:2014rta}
M.~Hogervorst, S.~Rychkov, and B.~C. van Rees, ``{Truncated conformal space
  approach in $d$ dimensions: A cheap alternative to lattice field theory?},''
  \href{http://dx.doi.org/10.1103/PhysRevD.91.025005}{{\em Phys. Rev.}
  {\bfseries D91} (2015) 025005},
\href{http://arxiv.org/abs/1409.1581}{{\ttfamily arXiv:1409.1581 [hep-th]}}.

\bibitem{Elias-Miro:2020qwz}
J.~Elias-Mir\'o and E.~Hardy, ``{Exploring Hamiltonian Truncation in
  $\bf{d=2+1}$},'' \href{http://dx.doi.org/10.1103/PhysRevD.102.065001}{{\em
  Phys. Rev. D} {\bfseries 102} no.~6, (2020) 065001},
  \href{http://arxiv.org/abs/2003.08405}{{\ttfamily arXiv:2003.08405
  [hep-th]}}.

\bibitem{Anand:2020qnp}
N.~Anand, E.~Katz, Z.~U. Khandker, and M.~T. Walters, ``{Nonperturbative
  dynamics of (2+1)d $\phi^4$-theory from Hamiltonian truncation},''
  \href{http://dx.doi.org/10.1007/JHEP05(2021)190}{{\em JHEP} {\bfseries 05}
  (2021) 190}, \href{http://arxiv.org/abs/2010.09730}{{\ttfamily
  arXiv:2010.09730 [hep-th]}}.

\bibitem{Klassen:1990dx}
T.~R. Klassen and E.~Melzer, ``{The Thermodynamics of purely elastic scattering
  theories and conformal perturbation theory},''
  \href{http://dx.doi.org/10.1016/0550-3213(91)90159-U}{{\em Nucl. Phys. B}
  {\bfseries 350} (1991) 635--689}.

\bibitem{Klassen:1991ze}
T.~R. Klassen and E.~Melzer, ``{Spectral flow between conformal field theories
  in (1+1) dimensions},''
\href{http://dx.doi.org/10.1016/0550-3213(92)90422-8}{{\em Nucl.Phys.}
  {\bfseries B370} (1992) 511--550}.

\bibitem{Giokas:2011ix}
P.~Giokas and G.~Watts, ``{The renormalisation group for the truncated
  conformal space approach on the cylinder},''
\href{http://arxiv.org/abs/1106.2448}{{\ttfamily arXiv:1106.2448 [hep-th]}}.

\bibitem{Rutter:2018aog}
D.~Rutter and B.~C. van Rees, ``{Counterterms in Truncated Conformal
  Perturbation Theory},'' \href{http://arxiv.org/abs/1803.05798}{{\ttfamily
  arXiv:1803.05798 [hep-th]}}.

\bibitem{Pappadopulo:2012jk}
D.~Pappadopulo, S.~Rychkov, J.~Espin, and R.~Rattazzi, ``{OPE Convergence in
  Conformal Field Theory},''
  \href{http://dx.doi.org/10.1103/PhysRevD.86.105043}{{\em Phys. Rev. D}
  {\bfseries 86} (2012) 105043},
  \href{http://arxiv.org/abs/1208.6449}{{\ttfamily arXiv:1208.6449 [hep-th]}}.

\bibitem{saleur1987two}
H.~Saleur and C.~Itzykson, ``Two-dimensional field theories close to
  criticality,'' {\em Journal of statistical physics} {\bfseries 48} no.~3,
  (1987) 449--475.

\bibitem{Belavin:1984vu}
A.~A. Belavin, A.~M. Polyakov, and A.~B. Zamolodchikov, ``{Infinite Conformal
  Symmetry in Two-Dimensional Quantum Field Theory},''
  \href{http://dx.doi.org/10.1016/0550-3213(84)90052-X}{{\em Nucl. Phys. B}
  {\bfseries 241} (1984) 333--380}.

\bibitem{Mussardo:2020rxh}
G.~Mussardo, {\em {Statistical Field Theory}}.
\newblock Oxford Graduate Texts. Oxford University Press, 3, 2020.

\bibitem{Zamolodchikov:1991vx}
A.~B. Zamolodchikov, ``{From tricritical Ising to critical Ising by
  thermodynamic Bethe ansatz},''
  \href{http://dx.doi.org/10.1016/0550-3213(91)90423-U}{{\em Nucl. Phys. B}
  {\bfseries 358} (1991) 524--546}.

\bibitem{Fendley:1993xa}
P.~Fendley, H.~Saleur, and A.~B. Zamolodchikov, ``{Massless flows, 2. The Exact
  S matrix approach},'' \href{http://dx.doi.org/10.1142/S0217751X93002277}{{\em
  Int. J. Mod. Phys. A} {\bfseries 8} (1993) 5751--5778},
  \href{http://arxiv.org/abs/hep-th/9304051}{{\ttfamily arXiv:hep-th/9304051}}.

\bibitem{Dotsenko:1984ad}
V.~S. Dotsenko and V.~A. Fateev, ``{Four Point Correlation Functions and the
  Operator Algebra in the Two-Dimensional Conformal Invariant Theories with the
  Central Charge c \ensuremath{<} 1},''
  \href{http://dx.doi.org/10.1016/S0550-3213(85)80004-3}{{\em Nucl. Phys. B}
  {\bfseries 251} (1985) 691--734}.

\bibitem{Dotsenko:1984nm}
V.~S. Dotsenko and V.~A. Fateev, ``{Conformal Algebra and Multipoint
  Correlation Functions in Two-Dimensional Statistical Models},''
  \href{http://dx.doi.org/10.1016/0550-3213(84)90269-4}{{\em Nucl. Phys. B}
  {\bfseries 240} (1984) 312}.

\bibitem{Cohen:2021erm}
T.~Cohen, K.~Farnsworth, R.~Houtz, and M.~A. Luty, ``{Hamiltonian Truncation
  Effective Theory},'' \href{http://arxiv.org/abs/2110.08273}{{\ttfamily
  arXiv:2110.08273 [hep-th]}}.

\bibitem{Rattazzi:2008pe}
R.~Rattazzi, V.~S. Rychkov, E.~Tonni, and A.~Vichi, ``{Bounding scalar operator
  dimensions in 4D CFT},''
  \href{http://dx.doi.org/10.1088/1126-6708/2008/12/031}{{\em JHEP} {\bfseries
  12} (2008) 031}, \href{http://arxiv.org/abs/0807.0004}{{\ttfamily
  arXiv:0807.0004 [hep-th]}}.

\bibitem{Poland:2018epd}
D.~Poland, S.~Rychkov, and A.~Vichi, ``{The Conformal Bootstrap: Theory,
  Numerical Techniques, and Applications},''
  \href{http://dx.doi.org/10.1103/RevModPhys.91.015002}{{\em Rev. Mod. Phys.}
  {\bfseries 91} (2019) 015002},
  \href{http://arxiv.org/abs/1805.04405}{{\ttfamily arXiv:1805.04405
  [hep-th]}}.

\bibitem{El-Showk:2012cjh}
S.~El-Showk, M.~F. Paulos, D.~Poland, S.~Rychkov, D.~Simmons-Duffin, and
  A.~Vichi, ``{Solving the 3D Ising Model with the Conformal Bootstrap},''
  \href{http://dx.doi.org/10.1103/PhysRevD.86.025022}{{\em Phys. Rev. D}
  {\bfseries 86} (2012) 025022},
  \href{http://arxiv.org/abs/1203.6064}{{\ttfamily arXiv:1203.6064 [hep-th]}}.

\bibitem{El-Showk:2014dwa}
S.~El-Showk, M.~F. Paulos, D.~Poland, S.~Rychkov, D.~Simmons-Duffin, and
  A.~Vichi, ``{Solving the 3d Ising Model with the Conformal Bootstrap II.
  c-Minimization and Precise Critical Exponents},''
  \href{http://dx.doi.org/10.1007/s10955-014-1042-7}{{\em J. Stat. Phys.}
  {\bfseries 157} (2014) 869}, \href{http://arxiv.org/abs/1403.4545}{{\ttfamily
  arXiv:1403.4545 [hep-th]}}.

\bibitem{Simmons-Duffin:2016wlq}
D.~Simmons-Duffin, ``{The Lightcone Bootstrap and the Spectrum of the 3d Ising
  CFT},'' \href{http://dx.doi.org/10.1007/JHEP03(2017)086}{{\em JHEP}
  {\bfseries 03} (2017) 086}, \href{http://arxiv.org/abs/1612.08471}{{\ttfamily
  arXiv:1612.08471 [hep-th]}}.

\bibitem{Chester:2019ifh}
S.~M. Chester, W.~Landry, J.~Liu, D.~Poland, D.~Simmons-Duffin, N.~Su, and
  A.~Vichi, ``{Carving out OPE space and precise $O(2)$ model critical
  exponents},'' \href{http://dx.doi.org/10.1007/JHEP06(2020)142}{{\em JHEP}
  {\bfseries 06} (2020) 142}, \href{http://arxiv.org/abs/1912.03324}{{\ttfamily
  arXiv:1912.03324 [hep-th]}}.

\bibitem{Liu:2020tpf}
J.~Liu, D.~Meltzer, D.~Poland, and D.~Simmons-Duffin, ``{The Lorentzian
  inversion formula and the spectrum of the 3d O(2) CFT},''
  \href{http://dx.doi.org/10.1007/JHEP09(2020)115}{{\em JHEP} {\bfseries 09}
  (2020) 115}, \href{http://arxiv.org/abs/2007.07914}{{\ttfamily
  arXiv:2007.07914 [hep-th]}}. [Erratum: JHEP 01, 206 (2021)].

\bibitem{Rychkov:2014eea}
S.~Rychkov and L.~G. Vitale, ``{Hamiltonian truncation study of the $?^4$
  theory in two dimensions},''
  \href{http://dx.doi.org/10.1103/PhysRevD.91.085011}{{\em Phys. Rev.}
  {\bfseries D91} (2015) 085011},
\href{http://arxiv.org/abs/1412.3460}{{\ttfamily arXiv:1412.3460 [hep-th]}}.

\bibitem{Rychkov:2015vap}
S.~Rychkov and L.~G. Vitale, ``{Hamiltonian truncation study of the $\phi^4$
  theory in two dimensions. II. The $\mathbb Z_2$ -broken phase and the Chang
  duality},'' \href{http://dx.doi.org/10.1103/PhysRevD.93.065014}{{\em Phys.
  Rev.} {\bfseries D93} no.~6, (2016) 065014},
\href{http://arxiv.org/abs/1512.00493}{{\ttfamily arXiv:1512.00493 [hep-th]}}.

\bibitem{Elias-Miro:2015bqk}
J.~Elias-Miro, M.~Montull, and M.~Riembau, ``{The renormalized Hamiltonian
  truncation method in the large $E_T$ expansion},''
  \href{http://dx.doi.org/10.1007/JHEP04(2016)144}{{\em JHEP} {\bfseries 04}
  (2016) 144},
\href{http://arxiv.org/abs/1512.05746}{{\ttfamily arXiv:1512.05746 [hep-th]}}.

\bibitem{Elias-Miro:2017xxf}
J.~Elias-Miro, S.~Rychkov, and L.~G. Vitale, ``{High-Precision Calculations in
  Strongly Coupled Quantum Field Theory with Next-to-Leading-Order Renormalized
  Hamiltonian Truncation},''
\href{http://arxiv.org/abs/1706.06121}{{\ttfamily arXiv:1706.06121 [hep-th]}}.

\bibitem{Elias-Miro:2017tup}
J.~Elias-Miro, S.~Rychkov, and L.~G. Vitale, ``{NLO Renormalization in the
  Hamiltonian Truncation},''
  \href{http://dx.doi.org/10.1103/PhysRevD.96.065024}{{\em Phys. Rev.}
  {\bfseries D96} no.~6, (2017) 065024},
\href{http://arxiv.org/abs/1706.09929}{{\ttfamily arXiv:1706.09929 [hep-th]}}.

\end{thebibliography}\endgroup
\bibliographystyle{utphys}

\end{document}